%% file: AliFemto2.tex
\documentclass[aps,prd,twocolumn,groupedaddress,superscriptaddress,altaffilletter,nobibnotes,nofootinbib,showpacs,10pt]{revtex4-1}

\usepackage{preprintcover}
\PreprintIdNumber{CERN-PH-EP-2010-083}
\PreprintCoverPaperTitle{Femtoscopy of \pp collisions at $\sqrt{s}=0.9$ and $7$~TeV
  at the LHC with two-pion Bose-Einstein correlations}
\PreprintCoverAbstract{We report on the high statistics two-pion correlation functions from 
  \pp collisions at $\sqrt{s}=0.9$~TeV and $\sqrt{s}=7$~TeV,
  measured by the ALICE experiment at the Large Hadron Collider. The
  correlation functions as well as the extracted source radii scale
  with event multiplicity and pair momentum. When analyzed in the same
  multiplicity and pair transverse momentum range, the correlation
  is similar at the two collision energies. A three-dimensional
  femtoscopic analysis shows an increase of the emission zone with
  increasing event multiplicity as well as decreasing homogeneity
  lengths with increasing transverse momentum. The latter trend gets
  more pronounced as multiplicity increases. This suggests the
  development of space-momentum correlations, at least for collisions
  producing a high multiplicity of particles. We consider these trends
  in the context of previous femtoscopic studies in high-energy hadron
  and heavy-ion collisions, and discuss possible underlying physics 
  mechanisms. Detailed analysis of the correlation reveals an
  exponential shape in the outward and longitudinal directions, while
  the sideward remains a Gaussian. This is interpreted as a result of a
  significant contribution of strongly decaying resonances to the
  emission region shape. Significant non-femtoscopic correlations are
  observed, and are argued to be the consequence of
  ``mini-jet''-like structures extending to low $p_{\rm T}$. They are
  well reproduced by the Monte-Carlo generators and seen also in
  $\pi^{+}\pi^{-}$ correlations. 
}
\PreprintJournalName{PRD}

\usepackage{graphicx,type1cm,eso-pic,color}
\usepackage{grffile}
\usepackage{epsfig}
\usepackage{graphicx}% Include figure files
\usepackage{fancyhdr}
\usepackage{pslatex}   
\usepackage{color}
\usepackage{amsmath, amsthm, amssymb} 
\usepackage{endnotes}

\definecolor{grey}{rgb}{0.3,0.3,0.3}
\definecolor{TBD}{rgb}{0.0,0.0,0.5}
\newcommand {\pp} {\mbox{$pp$}~}

\newcommand {\kt} {\mbox{$k_{\rm T}$}~}
\newcommand {\pt} {\mbox{$p_{\rm T}$}~}

\begin{document}
%\linenumbers

\newcommand{\inst}[1]{$^{#1}$}
%%%%%%%
\renewcommand{\and}{, }  %%%% this is somehow screwed up.  you might have to comment this command out, try pdflatex, then put it back, try pdflatex twice, etc...

\title {Femtoscopy of \pp collisions at $\sqrt{s}=0.9$ and $7$~TeV
  at the LHC with two-pion Bose-Einstein correlations}  
%\input{authors2.tex}
%\input{institutes2.tex}
\input{hbtpp-authors-revtex.tex}
%%% Do not change the next lines!
%\collaboration{The ALICE Collaboration%
%         \thanks{See Appendix~\ref{app:collab} for the list of collaboration 
%                      members}}
%\author{The ALICE Collaboration}      % appears on left page headers, do not change
%\author{K.~Aamodt {\it et al.} (ALICE Collaboration)}\thanks{Full author list given at the end of the article.}
 
\date{\today}

\begin{abstract}
  We report on the high statistics two-pion correlation functions from 
  \pp collisions at $\sqrt{s}=0.9$~TeV and $\sqrt{s}=7$~TeV,
  measured by the ALICE experiment at the Large Hadron Collider. The
  correlation functions as well as the extracted source radii scale
  with event multiplicity and pair momentum. When analyzed in the same
  multiplicity and pair transverse momentum range, the correlation
  is similar at the two collision energies. A three-dimensional
  femtoscopic analysis shows an increase of the emission zone with
  increasing event multiplicity as well as decreasing homogeneity
  lengths with increasing transverse momentum. The latter trend gets
  more pronounced as multiplicity increases. This suggests the
  development of space-momentum correlations, at least for collisions
  producing a high multiplicity of particles. We consider these trends
  in the context of previous femtoscopic studies in high-energy hadron
  and heavy-ion collisions, and discuss possible underlying physics 
  mechanisms. Detailed analysis of the correlation reveals an
  exponential shape in the outward and longitudinal directions, while
  the sideward remains a Gaussian. This is interpreted as a result of a
  significant contribution of strongly decaying resonances to the
  emission region shape. Significant non-femtoscopic correlations are
  observed, and are argued to be the consequence of
  ``mini-jet''-like structures extending to low $p_{\rm T}$. They are
  well reproduced by the Monte-Carlo generators and seen also in
  $\pi^{+}\pi^{-}$ correlations. 

\keywords{proton collisions, HBT, femtoscopy, intensity interferometry, LHC}
\end{abstract}
\pacs{25.75.-q, 25.75.Gz, 25.70.Pq}

\maketitle
 
%%%%%%%%%%%%%%%%%%%%%%%%%%%%%%%% Motivation for LHC and for measuring space scales of system

\section{Introduction}
\label{sec:intro}

Proton-proton collisions at $\sqrt{s}=0.9$~TeV and
$\sqrt{s}=7$~TeV have  been recorded by A Large Ion
Collider Experiment (ALICE) at the  
Large Hadron Collider (LHC) at CERN in 2010.  
These collisions provide a unique opportunity to probe Quantum
Chromodynamics (QCD) in the new energy regime. The
distinguishing feature of QCD is the mechanism of color confinement,
the physics of which is not fully understood, due, in part, to its
theoretical intractability~\cite{Fodor:2001pe}.  The confinement
mechanism has a physical scale of the order of the proton radius and
is especially important at low momentum. The study presented in this
work aims to measure the space-time extent of the source on this scale.

%%%%%%%%%%%%%%%%%%%%%%%%% HBT paragraph

Two-pion correlations at low relative momentum were first shown to be
sensitive to the spatial scale of the emitting source in $\bar{p}+p$
collisions by G.~Goldhaber, S.~Goldhaber, W.~Lee and A.~Pais 50 years
ago~\cite{Goldhaber:1960sf}.   Since then, they were studied in
$e^{+}+e^-$~\cite{Kittel:2001zw}, hadron- and
lepton-hadron~\cite{Alexander:2003ug}, and heavy ion~\cite{Lisa:2005dd}
collisions. Especially in the latter case, two-particle femtoscopy has
been developed into a precision tool to probe the
dynamically-generated geometry structure of the emitting system. In
particular, a sharp phase transition between the color-deconfined and
confined states was precluded by the observation of short timescales,
and femtoscopic measurement of bulk collective flow proved that a
strongly self-interacting system was created in the
collision~\cite{Hardtke:1999vf,Pratt:2009hu}. 

Femtoscopy in heavy-ion collisions is believed to be understood in
some detail, see e.g.~\cite{Lisa:2005dd}. The spatial scales grow
naturally with the multiplicity of the event. Strong hydrodynamical
collective flow in the longitudinal and transverse directions is
revealed by dynamical dependencies of femtoscopic scales. 
The main puzzling aspect of the data is the relative
energy independence of the results of the measurements. 

To some extent, Bose-Einstein correlations in particle physics were
initially of interest only as a source of systematic uncertainty in
the determination of the $W$ boson mass~\cite{Lonnblad:1998xx}. But
overviews~\cite{Kittel:2001zw,Alexander:2003ug,Chajecki:2009zg} of
femtoscopic measurements in hadron- and lepton-induced collisions
reveal systematics surprisingly similar to those mentioned above for
heavy-ion collisions. Moreover, in the first direct comparison of
femtoscopy in heavy-ion collisions at Relativistic Heavy-Ion Collider
(RHIC) and proton collisions in the same apparatus an
essentially identical multiplicity- and momentum-dependence was reported in the
two systems~\cite{STAR:2010bw}. However, the multiplicities at which the 
femtoscopic measurement in \pp collisions at RHIC was made were still
significantly smaller than those in even the most peripheral
heavy-ion collisions. In this work we are, for the first time, able to
compare femtoscopic radii measured in \pp and heavy-ion collisions at
comparable event multiplicities. At these multiplicities the observed
correlations may be influenced by jets~\cite{Paic:2005cx} while other
studies suggest that a system behaving collectively may be
created~\cite{Bozek:2009dt}.
  
In our previous work~\cite{Aamodt:2010jj} we reported that a
multiplicity integrated measurement does not show any pair momentum
dependence of the $R_{inv}$ radius measured in the Pair Rest Frame
(PRF). Similar analysis from the CMS
collaboration~\cite{Khachatryan:2010un} also mentions that no momentum
dependence was observed. However the analysis in two multiplicity
ranges suggested that momentum dependence may change with
multiplicity, although any strong conclusions were precluded by
limited statistics. In this work we explore this dependence by using
high statistics data and more multiplicity ranges. It enabled us to
perform the three-dimensional analysis in the Longitudinally Co-Moving
System (LCMS), where the pair momentum along the beam vanishes. 

The paper is organized as follows: in Section~\ref{sec:alicedata} we
describe the ALICE experimental setup and data taking conditions for
the sample used in this work. In Section~\ref{sec:correlation} we
present the correlation measurement and characterize the correlation
functions themselves. In Section~\ref{sec:3dradii} we show the main
results of this work: the three-dimensional radii extracted from the
data. We discuss various observed features and compare the
results to other experiments. 
In Section~\ref{sec:1dfit} we
show, for completeness, the one-dimensional $R_{inv}$ analysis. Finally
in Section~\ref{sec:summary} we summarize our results. All the
numerical values can be obtained from the Durham Reaction
Database~\cite{Durham:2010aa}.  

%%%%%%%%%%%%%%%%%%%%%%%%%%%%%%%%%%% ALICE datataking

\section{ALICE data taking}
\label{sec:alicedata}

In this study we report on the analysis of \pp collisions recorded
by the ALICE experiment during the 2010 run of the LHC. Approximately
8 million events, triggered by a minimum bias 
trigger at the injection energy of $\sqrt{s} = 0.9$~TeV, and 100
million events with similar trigger at the maximum LHC energy to date,
$\sqrt{s} = 7$~TeV, were analyzed in this work. 

The ALICE Time Projection Chamber (TPC)~\cite{Aamodt:2008zz} was used to
record charged particle tracks as they left ionization trails in the
${\rm Ne-CO_2}$ gas.  The ionization drifts up to 2.5~m from the
central electrode to the end-caps to be measured on 159~padrows, which
are grouped into 18 sectors; the position at which the track crossed
the padrow was determined with resolutions of 2~mm and 3~mm in the
drift and transverse directions, respectively. The 
momentum resolution is $\sim 1\%$ for pions with $p_{\rm
  T}=0.5$~GeV/$c$. The ALICE Inner Tracking   System (ITS) was also
used for tracking. It consists of six silicon layers, two innermost
Silicon Pixel Detector (SPD) layers, two Silicon Drift Detector (SDD)
layers, and two outer Silicon Strip Detector (SSD) layers, which
provide up to six space 
points for each track. The tracks used in this analysis were
reconstructed using the information from both the TPC and the 
ITS, such tracks were also used to reconstruct the primary vertex of 
the collision. For details of this procedure and its efficiency
see~\cite{Aamodt:2010pp}. The forward scintillator detectors VZERO are
placed along the beam line at $+3$~m and $-0.9$~m from the nominal
interaction point. They cover a region $2.8 < \eta < 5.1$ and
$-3.7 < \eta < -1.7$ respectively. They were used in the minimum
bias trigger and their timing signal was used to reject the beam--gas
and beam-halo collisions. 

The minimum bias trigger required a signal in either of the two VZERO
counters or one of the two inner layers of the Silicon Pixel Detector
(SPD). Within this sample, we selected events based on the 
measured charged-particle multiplicity within the pseudorapidity range
$|\eta|<1.2$. Events were required to have a primary vertex
 within 1~mm of the beam line,
and 10~cm of the center of the 5~m-long TPC. This provides
almost uniform acceptance for particles with $|\eta|<1$ for all events
in the sample. It decreases for $1.0<|\eta|<1.2$. In
addition, we require events to have at least one charged particle
reconstructed within $|\eta|<1.2$.  

The minimum number of clusters associated to the track in the TPC is
70 (out of the maximum of 159) and 2 in the ITS (out of the maximum of
6). The quality of the track is determined by the
$\chi^2/N$ value for the Kalman fit to the reconstructed position of
the TPC clusters ($N$ is the number of clusters attached to the
track); the track is rejected if the value is larger than 
4.0 (2 degrees of freedom per cluster). Tracks with $|\eta|<1.2$ are
taken for the analysis. The $p_{\rm T}$ of accepted particles has a
lower limit of $0.13$~GeV/$c$, because tracks with lower  $p_{\rm T}$
do not cross enough padrows in the TPC. The efficiency of particle
reconstruction is about 50\%  at this lowest limit and then quickly
increases and 
reaches a stable value of approximately 80\% for $p_{\rm
  T}>0.2$~GeV/$c$. In order to reduce the number of secondary
particles in our sample, we require the track to project back to the
primary interaction vertex within $0.018+0.035 p_{\rm T}^{-1.01}$~cm
in the transverse plane and $0.3$~cm in the longitudinal direction
(so-called Distance of Closest Approach or DCA
selection). 

ALICE provides an excellent particle identification capability, through
the combination of the measurement of the specific ionization
(${\rm d}E/{\rm d}x$) in the TPC and the ITS and the timing signals in the ALICE
Time Of Flight (TOF). In the momentum range  covered here
($0.13$~GeV/$c$ to $0.7$~GeV/$c$) pions constitute the majority of
particles. We use only the TPC measurement for Particle IDentification
(PID) in this work, as the other detectors offer significant
improvement at higher $p_{\rm T}$ than used here. This PID procedure
results in a small contamination of the pion sample by electrons at
$p_{\rm   T}<0.2$~GeV/$c$ and kaons at $p_{\rm
  T}>0.65$~GeV/$c$. Allowing other particles into our sample has only
a minor effect of lowering the strength of the correlation (the
$\lambda$ parameter), while it does not affect the femtoscopic
radius, so we do not correct for it explicitly. The amount of electron
contamination is less than 5\%.

%%%%%%%%%%%%%%%%%%%%%%%%%%%%%%%%%%% The experiment
\section{Correlation function measurement}
\label{sec:correlation}

Experimentally, the two-particle correlation function is defined as
the ratio $C\left({\bf q}\right)=A\left({\bf q}\right)/B\left({\bf
    q}\right)$, where  $A\left({\bf q}\right)$ is the measured  
two-pion distribution of pair momentum difference ${\bf q}={\bf
  p}_2-{\bf p}_1$, and $B\left({\bf q}\right)$ is a similar
distribution formed by using pairs of particles from different
events~\cite{Kopylov:1974th}.

\begin{table}[tb]
\caption{Multiplicity selection for the analyzed sample. Uncorrected 
  $N_{ch}$ in $|\eta|<1.2$, $\left <{\rm d}N_{ch}/{\rm d}\eta \right >|_{N_{ch}\geq1}$
  (see text for the definition), number of events and number of
  identical pion pairs in each range are given. 
  \label{tab:mult}}
\begin{tabular}{lcccc}
  \hline
  \hline
  Bin & $N_{ch}$ & $\left <{\rm d}N_{ch}/{\rm d}\eta \right >|_{N_{ch}\geq1}$ & No. events
  $\times 10^6$
  & No. pairs $\times 10^6$ \\
  \hline
  \multicolumn{5}{c}{$\sqrt{s}=0.9$~TeV}\\
  \hline
  1 & 1--11   &  2.7 & 3.1   & 8.8 \\
  2 & 12--16  &  7.0 & 0.685 & 8.6 \\
  3 & 17--22  &  9.7 & 0.388 & 9.5 \\
  4 & 23--80  & 14.6 & 0.237 & 12.9 \\
  \hline
  \multicolumn{5}{c}{$\sqrt{s}=7$~TeV}\\
  \hline
  1 & 1--11   & 3.2   & 31.4  & 48.7  \\
  2 & 12--16  & 7.4   & 9.2   & 65.0  \\
  3 & 17--22  & 10.4  & 7.4   & 105.7 \\
  4 & 23--29  & 13.6  & 4.8   & 120.5 \\
  5 & 30--36  & 17.1  & 3.0   & 116.3 \\
  6 & 37--44  & 20.2  & 2.0   & 115.6 \\
  7 & 45--57  & 24.2  & 1.3   & 114.5 \\
  8 & 58--149 & 31.1  & 0.72  & 108.8 \\
  \hline
  \hline
\end{tabular}
\end{table}

The size of the data sample used for this analysis allowed for a highly
differential measurement. In order to address the physics topics
mentioned in the introduction, the analysis was performed
simultaneously as a function of the total event multiplicity $N_{ch}$
and pair transverse momentum  
$k_{\rm T}= \left | \vec{p}_{{\rm T},1} + \vec{p}_{{\rm T},2} \right
|/2$. 
For the multiplicity determination 
we 
counted the tracks
reconstructed simultaneously in the ITS and the TPC, plus the tracks
reconstructed only in the ITS in case the track was outside of the TPC 
$\eta$ acceptance. The total number of events accepted after applying
the selection criteria in the $\sqrt{s}=7$~TeV sample was $60 \times
10^6$ and in the $\sqrt{s}=0.9$~TeV sample it was $4.42 \times
10^6$. We divided the full multiplicity range into eight and four
ranges for the two energies respectively in such a way that the
like-charge pion pair multiplicity in each of them was
comparable. Table~\ref{tab:mult} gives (a) values for the range of raw
charged particle multiplicity
that was used to categorize the event, (b) the corresponding mean
charge particle density $\left<{\rm d}N_{ch}/{\rm d}\eta\right>$ as well as (c)
number of events and (d) the number of identical pion pairs in each
range. The femtoscopic measurement requires the events to have at
least one charged pion identified\footnote{In  
  fact the correlation signal is constructed from events having at
  least {\it two} same-charge pions (a pair). The one-pion events do
  contribute to the mixed background.}
and its momentum determined. 
We give the ${\rm d}N_{ch}/{\rm d}\eta$ values in Tab.~\ref{tab:mult} for this
event sample.
We denote this value as $\left < {\rm d}N_{ch}/{\rm d}\eta\right
>|_{N_{ch}\geq1}$; its typical uncertainty is 10\%. We note that for
the the lowest multiplicity this charged particle density is biased
towards higher values with respect to the full sample of inelastic
events. 

\begin{figure}[t!]
\centerline{
\includegraphics[width=0.40\textwidth]{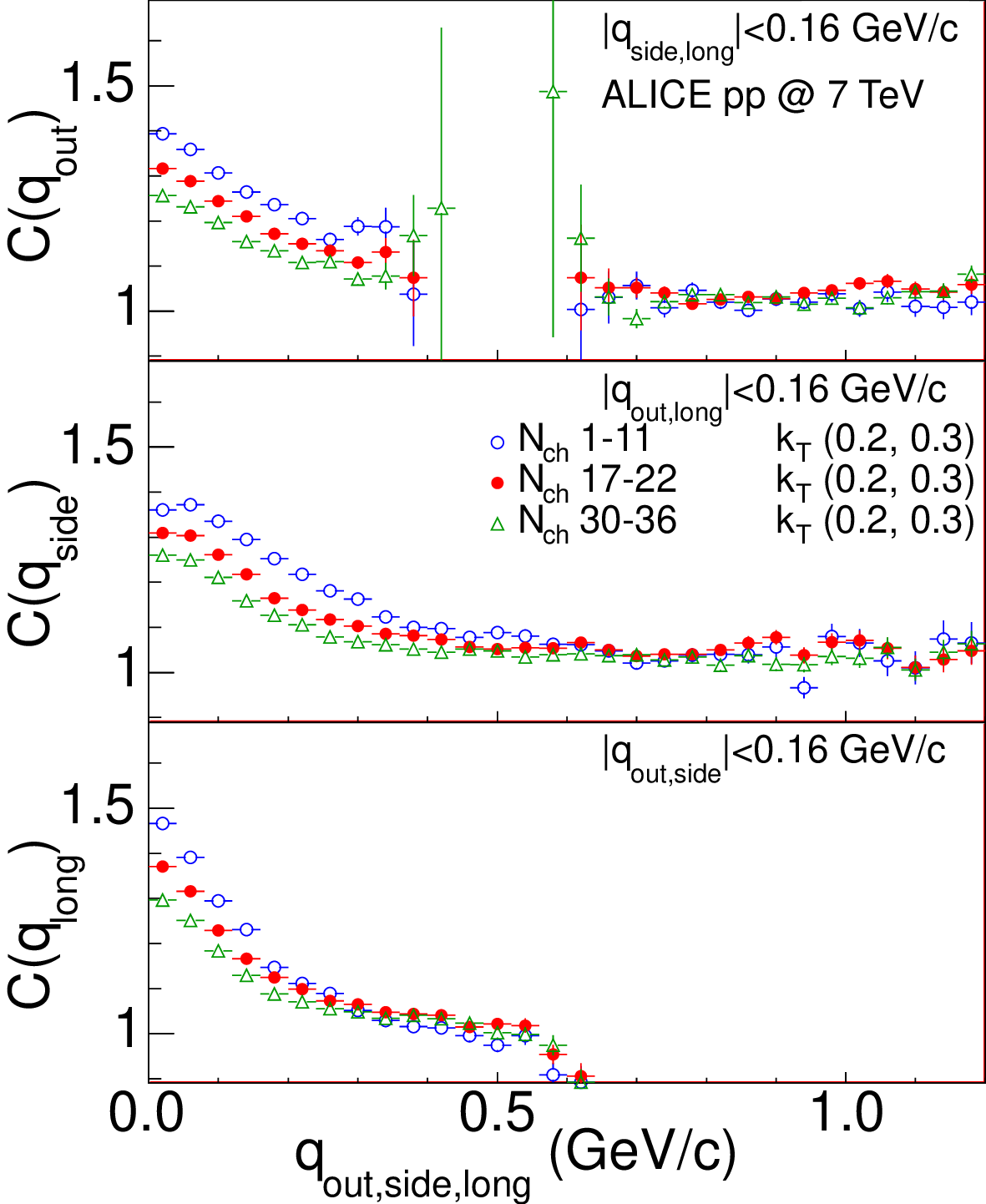}
}
\caption{\label{fig:projectionsct}
Projections of the 3D Cartesian representations of the correlation
functions onto the $q_{out}$, $q_{side}$, and $q_{long}$ axes for
pairs with $0.2 < k_{\rm T} < 0.3$~GeV/$c$, for three multiplicity
ranges. To project onto one $q$-component, the others are integrated 
over the range $0-0.16$~GeV/$c$. 
}
\end{figure}

The pair momentum $k_{\rm T}$ ranges used in the analysis were $(0.13,
0.2)$, $(0.2,0.3)$, $(0.3,0.4)$, $(0.4,0.5)$, $(0.5,0.6)$,
$(0.6,0.7)$~GeV/$c$.  

\subsection{Correlation function representations}
\label{sec:representations}

The correlations are measured as a function of pair relative momentum
four-vector {\bf q}. We deal with pions, so the masses of the
particles are fixed - in this case ${\bf q}$ reduces to a vector:
$\vec q$. The one-dimensional analysis is performed versus the magnitude
of the invariant momentum difference $q_{inv}=|\vec q|$, in PRF. The large
available statistics for this work allowed us to perform a detailed
analysis also for the 3D functions. In forming them, we calculate the
momentum difference in LCMS and decompose this ${\vec{q}_{LCMS}}$
according to the Bertsch--Pratt~\cite{Pratt:1986cc,Bertsch:1988db}
``out-side-long'' (sometimes indicated by $o$, $s$, and $l$
subscripts) parametrization. Here, $q_{long}$ is parallel to the beam,
$q_{out}$ is  parallel to the pair transverse momentum, and $q_{side}$
is perpendicular to $q_{long}$ and  $q_{out}$. If one wishes
to compare the radii measured in LCMS to $R_{inv}$ one needs to
multiply one of the transverse radii in LCMS (the one along the pair
transverse momentum) by the Lorentz $\gamma$ corresponding to the pair
transverse velocity, and then average the three radii. Therefore an
$R_{inv}$ constant with momentum is consistent with the radii in LCMS
decreasing with momentum. 
Figure~\ref{fig:projectionsct}
shows one-dimensional projections of the 3-dimensional correlation
function $C(q_{out},q_{side},q_{long})$ onto the $q_{out}$,
$q_{side}$, and $q_{long}$ axes, for $\pi^+$ pairs from one of the 
multiplicity/$k_{\rm T}$ ranges from the $\sqrt{s}=7$~TeV sample. The
function is normalized with a factor that is a result of the fit (the
details of the procedure are described in Sec.~\ref{sec:3dfits});
unity means no correlation.

\begin{figure}[t!]
\centerline{
\includegraphics[width=0.4\textwidth]{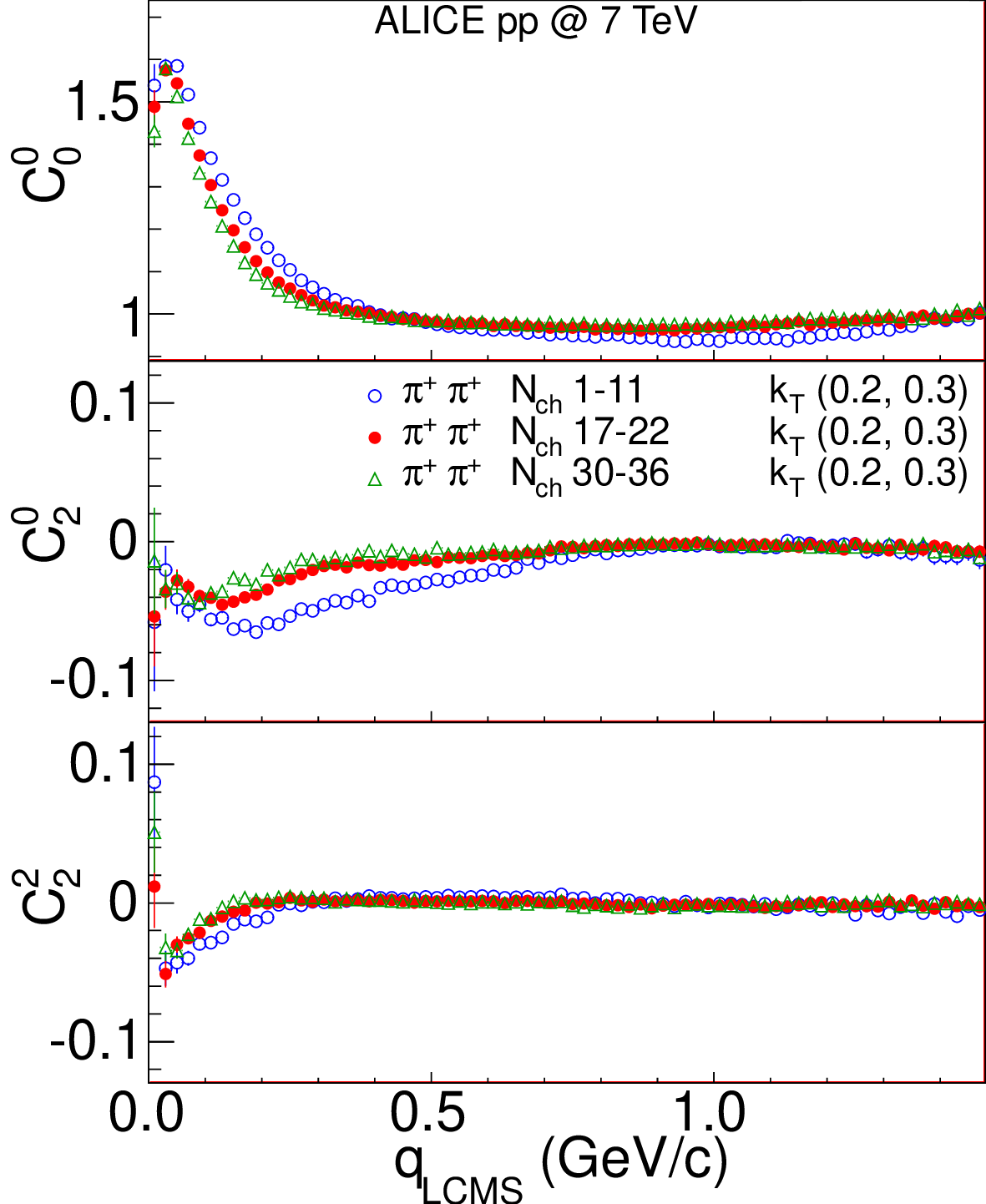}
}
\caption{\label{fig:projections}
Moments of the SH decomposition of the correlation
functions for pairs with $0.2 < k_{\rm T} < 0.3$~GeV/$c$, for three
multiplicity ranges.
}
\end{figure}

The 1-dimensional projections, shown in Fig.~\ref{fig:projectionsct},
present a limited view of the 3-dimensional structure of the
correlation function. It is increasingly common to represent
correlation functions in a harmonic
analysis~\cite{Brown:1997ku,Chajecki:2006hn,Kisiel:2009iw}; this
provides a more complete representation of the 3-dimensional structure
of the correlation, a better diagnostic of non-femtoscopic 
correlations~\cite{Chajecki:2006hn}, and a more direct relation to the
shape of the source~\cite{Danielewicz:2006hi}. The moments of  the
Spherical Harmonic (SH) decomposition are given by   

\begin{equation}
\label{eq:Alm}
A_l^m\left(|\vec q|\right) \equiv \frac{1}{\sqrt{4\pi}} \int d\phi
d(\cos\theta)  C\left(|\vec q|,\theta,\phi \right)
Y_l^m\left(\theta,\phi\right). 
\end{equation}
Here, the out-side-long space is mapped onto Euler angles in which
$q_{long}=|\vec q|\cos\theta$ and $q_{out}=|\vec
q|\sin\theta\cos\phi$. For pairs of identical particles in collider
experiments done with symmetrical beams, including the analysis in
this work, the odd $l$ and the imaginary and odd $m$ components for
even $l$ vanish. The first three non-vanishing moments, which capture
essentially all of the 3-dimensional structure, are then $C_0^0$,
$C_2^0$, and $C_2^2$. These are shown in
Fig.~\ref{fig:projections}. The components for $l\geq4$ represent the
fine details of the correlation structure and are not analyzed in this
work.  

The $C_0^0$ is the angle-averaged component. It captures the
general shape of the correlation. The width of the peak near $q=0$
is inversely proportional to the overall femtoscopic size of the
system. The $C_2^0$ component is the correlation weighed with the
$\cos^2(\theta)$. If it differs from 0, it signifies that the
longitudinal and transverse sizes of the emission region differ. The
$C_2^2$ is weighed with $\cos^2(\phi)$. If it differs from 0, it
signals that the outward and sideward sizes differ. The
correlation function is normalized to the number of pairs in the
background divided by the number of pairs in the signal.

\subsection{Measured correlations}

\begin{figure}[t!]
\centerline{
\includegraphics[width=0.4\textwidth]{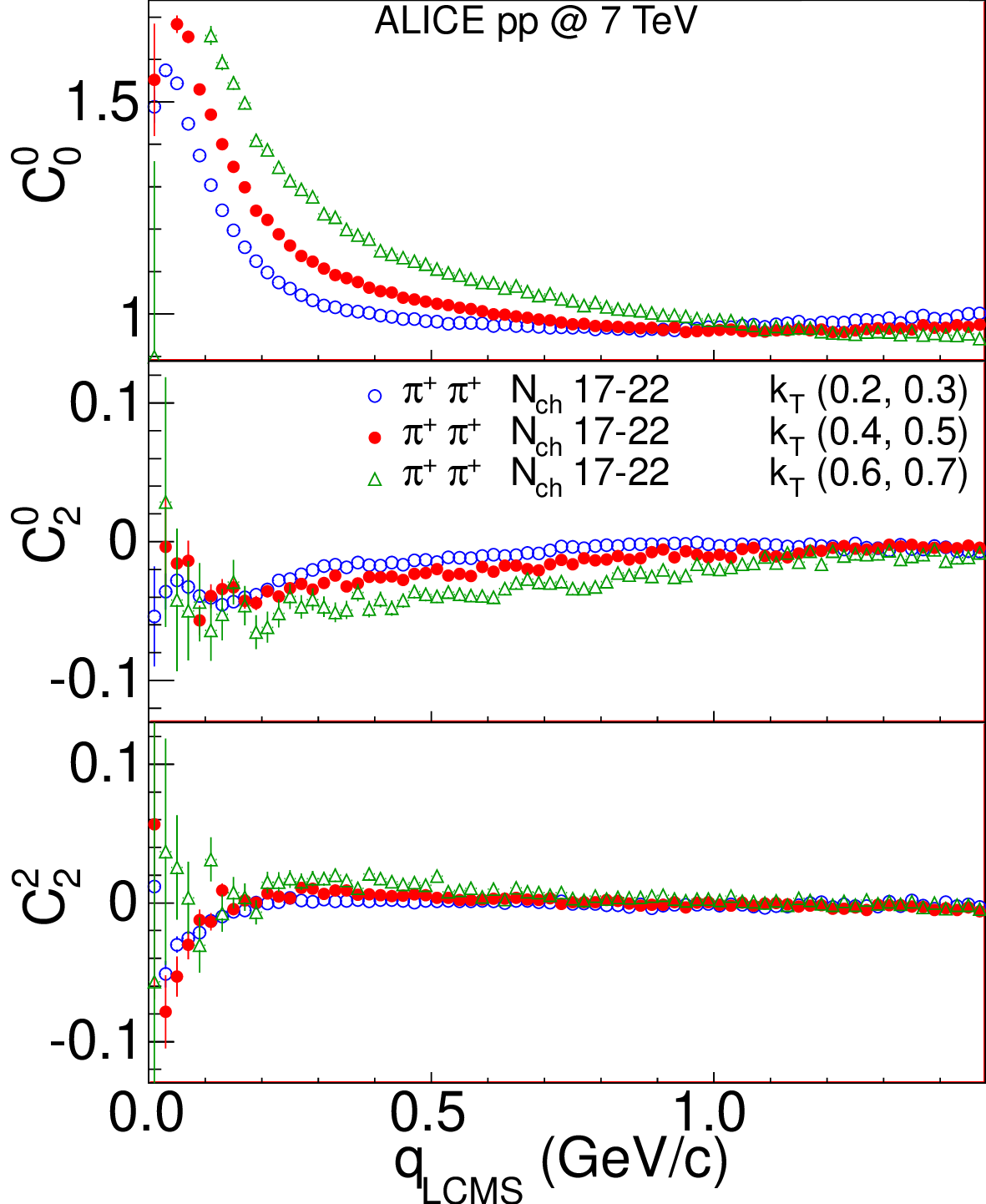}
}
\caption{\label{fig:ktdepcf}
  Moments of the SH decomposition of the correlation
  functions for events with $17 \leq N_{ch} \leq 22$, for three   $k_{\rm
    T}$ ranges.  
}
\end{figure}

In Figs.~\ref{fig:projectionsct} and~\ref{fig:projections} we show
selected correlations to  illustrate how they depend on
multiplicity. This is done for $k_{\rm T}$ of $(0.2,0.3)$~GeV/$c$; the
behavior in other $k_{\rm T}$ ranges and at the lower collision energy
is qualitatively the same. The narrowing of the correlation peak with
increasing multiplicity is apparent, corresponding to the increase of
the size of the emitting region. The behavior of the correlation
function at large $q$ is also changing, the low multiplicity baseline
is not flat, goes below 1.0 around $q=1$~GeV/$c$ and then rises again
at larger $q$, for higher multiplicities the background becomes
flatter at large $q$. In Cartesian representation shown in
Fig.~\ref{fig:projectionsct}, areas with no data points (acceptance
holes) are seen in  $q_{out}$ projection near $q=0.5$~GeV/$c$ and in
$q_{long}$ above $0.6$~GeV/c. Since $q_{long}$ is proportional to the
difference of longitudinal momenta, its value is limited due to $\eta$
acceptance. 
In the $out$ direction the hole appears due to a combination of lower \pt
cut-off and the selected \kt range. It can be simply understood as
follows: For the projection in the upper panel of
Fig.~\ref{fig:projectionsct}, we take the value of $q_{side}$ and
$q_{long}$ small. The value of $q_{side}$ is proportional to the
azimuthal angle difference, while $q_{long}$ is proportional to polar
angle difference. For $q_{side},q_{long}=0$, $q_{out}$ is simply
$p_{{\rm T},2} - p_{{\rm T},1}$ and \kt is $(p_{{\rm T},1} + p_{{\rm
    T},2})/2$, where \pt is no longer a 2-vector, but just a
scalar. The particles are either fully aligned (both $p_{\rm T}$'s are
positive or both are negative) or back-to-back (one \pt is positive,
the other negative). When we combine the lower \pt
cut-off $|p_{\rm T}|>0.13$~GeV/$c$ and the \kt selection $0.2 \leq
k_{\rm T} \leq 0.3$, it can be shown that some range of the $q_{out}$
values is excluded. This range will depend on the \kt selection.

\begin{figure}[t!]
\centerline{
\includegraphics[width=0.4\textwidth]{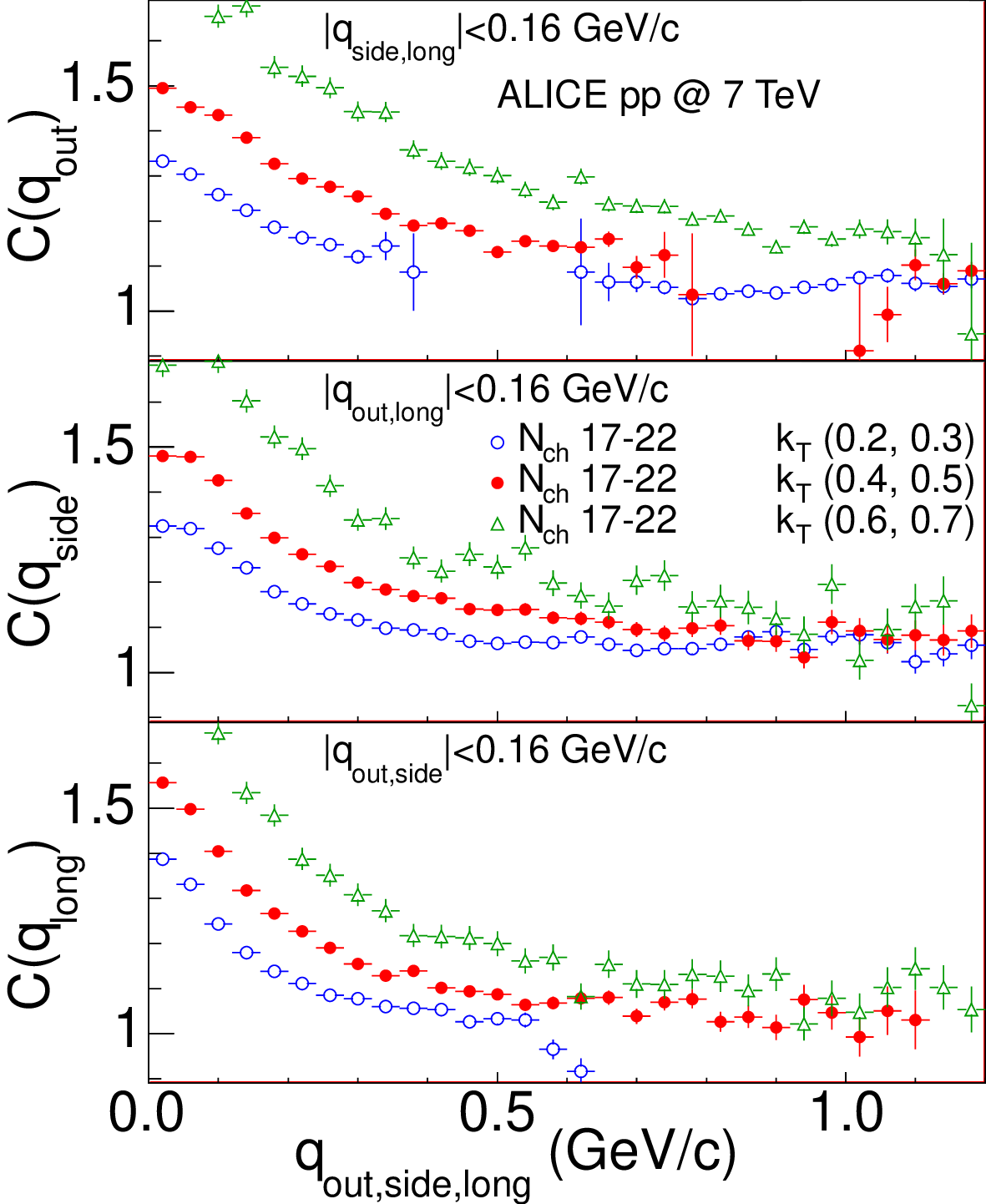}
}
\caption{\label{fig:ktdepcfct}
  Projections of the 3D Cartesian representations of the correlation
  functions onto the $q_{out}$, $q_{side}$, and $q_{long}$
  axes, for events with $17 \leq N_{ch} \leq 22$, for three
  $k_{\rm T}$ ranges. To project onto one $q$-component, the others
  are integrated over the range $0-0.16$~GeV/$c$.
}
\end{figure}

The $k_{\rm T}$ dependence of the correlation function is shown in
Figs.~\ref{fig:ktdepcf} and~\ref{fig:ktdepcfct}, for multiplicity 
$17 \leq N_{ch} \leq22$. The behavior in other multiplicity ranges and
at lower energy is qualitatively similar (except the lowest
multiplicity bin where the behavior is more complicated - see the
discussion of the extracted radii in Sec.~\ref{sec:3dfits} for
details). We see a strong change of 
the correlation with $k_{\rm T}$, with two apparent effects. At low
$k_{\rm T}$ the correlation appears to be dominated by the femtoscopic
effect at $q<0.3$~GeV/$c$, and is flat at larger $q$. As $k_{\rm T}$
grows, the femtoscopic peak  broadens (corresponding to a decrease in
size of the emitting region). In addition, a wide structure, extending up to $1.0$~GeV/$c$ in
$q$ for the highest $k_{\rm T}$ range, appears. We analyze this
structure in further detail later in this work. We also see that,
according to expectations, the acceptance holes in the $out$ and
$long$ region move as we change the \kt range.

\begin{figure}[t!]
\centerline{
\includegraphics[width=0.4\textwidth]{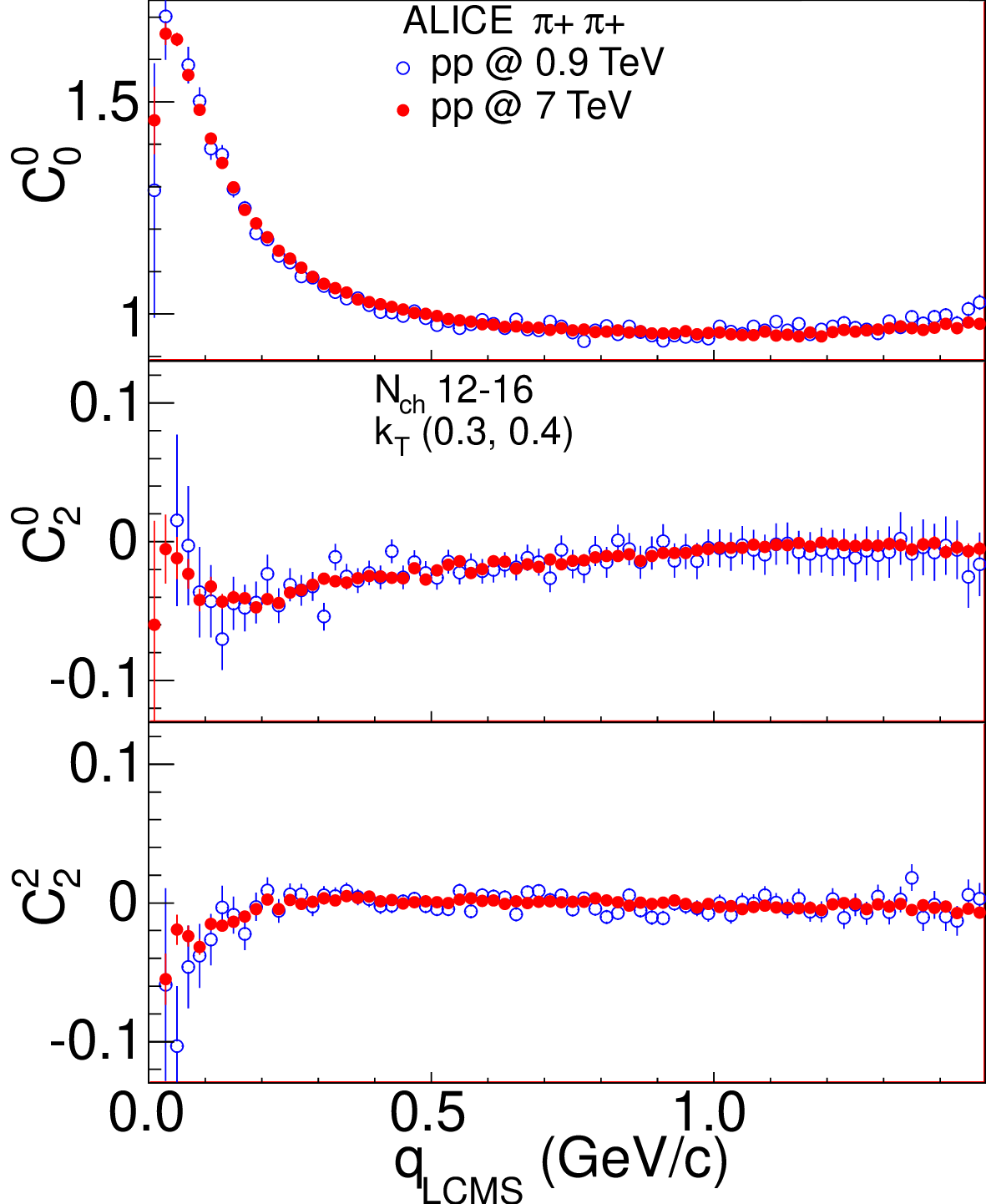}
} 
\caption{\label{fig:endepcf}
  Moments of the SH decomposition of the correlation functions for
  events with $12 \leq N_{ch} \leq 16$, pairs with $0.3 < k_{\rm T} <
  0.4$~GeV/$c$. Open symbols are for $\sqrt{s} =   0.9$~TeV
  collisions, closed symbols for $\sqrt{s} = 7$~TeV collisions.  
}
\end{figure}

Figure~\ref{fig:endepcf} shows the example of the correlation
function, for the same multiplicity/$k_{\rm T}$ range, for \pp collisions
at two collision energies. We note a similarity between the two
functions; the same is seen for other $k_{\rm T}$'s and overlapping
multiplicity ranges. The similarity is not trivial: changing the
multiplicity by 50\%, as seen in Fig.~\ref{fig:projections} or $k_{\rm
  T}$ by 30\% as seen in Fig.~\ref{fig:ktdepcf} has a stronger
influence on the correlation function than changing the collision
energy by an order of magnitude. We conclude that the main scaling
variables for the correlation function are global event multiplicity
and transverse momentum of the pair; the dependence on collision
energy is small. The energy independence of the emission region size
is the first important physics result of this work. We emphasize that
it can be already drawn from the analysis of the correlation functions
themselves, but we will also perform more qualitative checks and
discussions when we report the fitted emission region sizes in
Section~\ref{sec:fits}. 

\begin{figure}[t!]
\centerline{
\includegraphics[width=0.4\textwidth]{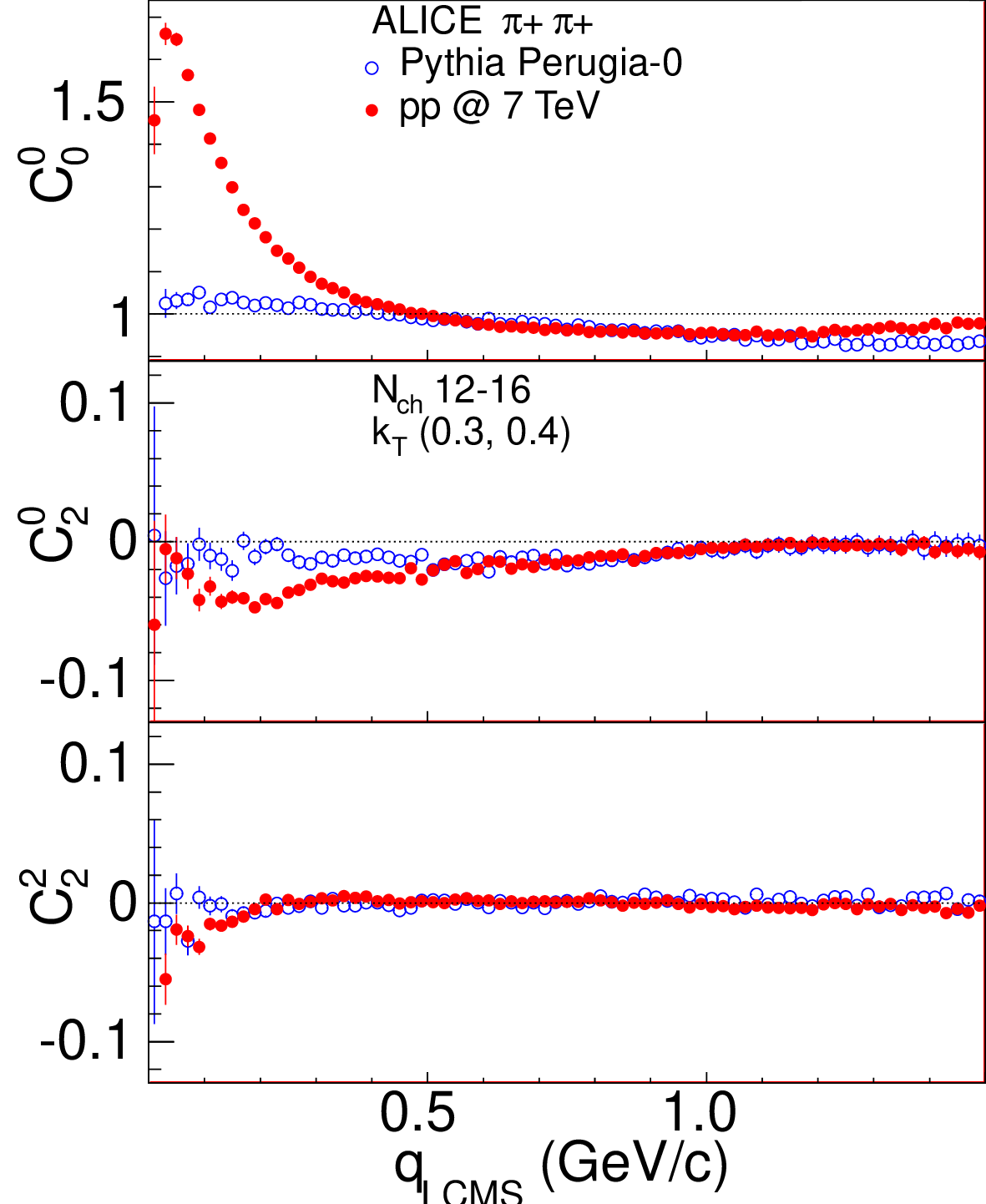}
}
\caption{\label{fig:mcvsdatacf}
  Moments of the SH decomposition of the correlation
  functions for events with $12 \leq N_{ch} \leq 16$, pairs with
  $0.3 < k_{\rm T} < 0.4$~GeV/$c$. Open symbols are {\sc PYTHIA} MC
  simulations (Perugia-0 tune), closed symbols are ALICE data from
  $\sqrt{s} = 7$~TeV collisions.  
}
\end{figure}
  
\subsection{Non-femtoscopic correlation structures}
\label{sec:nonfemto}

In Fig.~\ref{fig:ktdepcf} we noted the appearance of long-range
structures in the correlation functions for large $k_{\rm T}$. If these
were of femtoscopic origin, they would correspond to an unusually small
emission region size of~$0.2$~fm. We reported the observation of these
structures in our previous analysis~\cite{Aamodt:2010jj} at
$\sqrt{s}=0.9$~TeV, where they 
were interpreted as non-femtoscopic correlations coming from
``mini-jet'' like structures at $p_{\rm T}<1$~GeV/$c$. Here we further analyze
this hypothesis. In Fig.~\ref{fig:mcvsdatacf} we show the comparison
of the correlation function at multiplicity $12\leq N_{ch} \leq16$ in
an intermediate $k_{\rm T}$ range, where the long-range correlations are
apparent, to the Monte-Carlo (MC) calculation. The simulation used the
{\sc pythia} generator~\cite{Sjostrand:2006za}, Perugia-0
tune~\cite{Skands:2009zm} as input and was propagated through the full
simulation of the ALICE detector~\cite{Aamodt:2008zz}. Then it was
reconstructed and analyzed in exactly the same way as our real data,
using the same multiplicity and $k_{\rm T}$ ranges. The MC
calculation does not include the wave-function symmetrization for
identical particles; hence, the absence of the femtoscopic peak at low
$q$ is expected. In the angle-averaged $C_0^0$ component a significant
correlation structure is seen, up to $1$~GeV/$c$, with a slope similar
to the data outside of the peak at low $q$. Similarly, in the $C_2^0$
component a weak and wide correlation dip is seen around
$q=0.5$~GeV/$c$, which is also seen in the data. In MC, the
correlation in $C_2^0$ disappears at lower $q$, while for the data it
extends to much lower $q$, exactly where the femtoscopic peak is
expected and  seen in $C_0^0$. Our hypothesis is that both the
long-range peak in $C_0^0$ and the dip in $C_2^0$ are 
of a ``mini-jet'' origin. They need to be taken into account when
fitting the correlation function from data, so that the femtoscopic
peak can be properly extracted and characterized. The calculations
was also carried out with a second Monte-Carlo, the {\sc phojet}
model~\cite{Engel:1994vs,Engel:1995yda}, and gave similar
results. The differences between the two models are reflected in the
systematic error.  

\begin{figure*}[]
\centerline{
\includegraphics[width=0.80\textwidth]{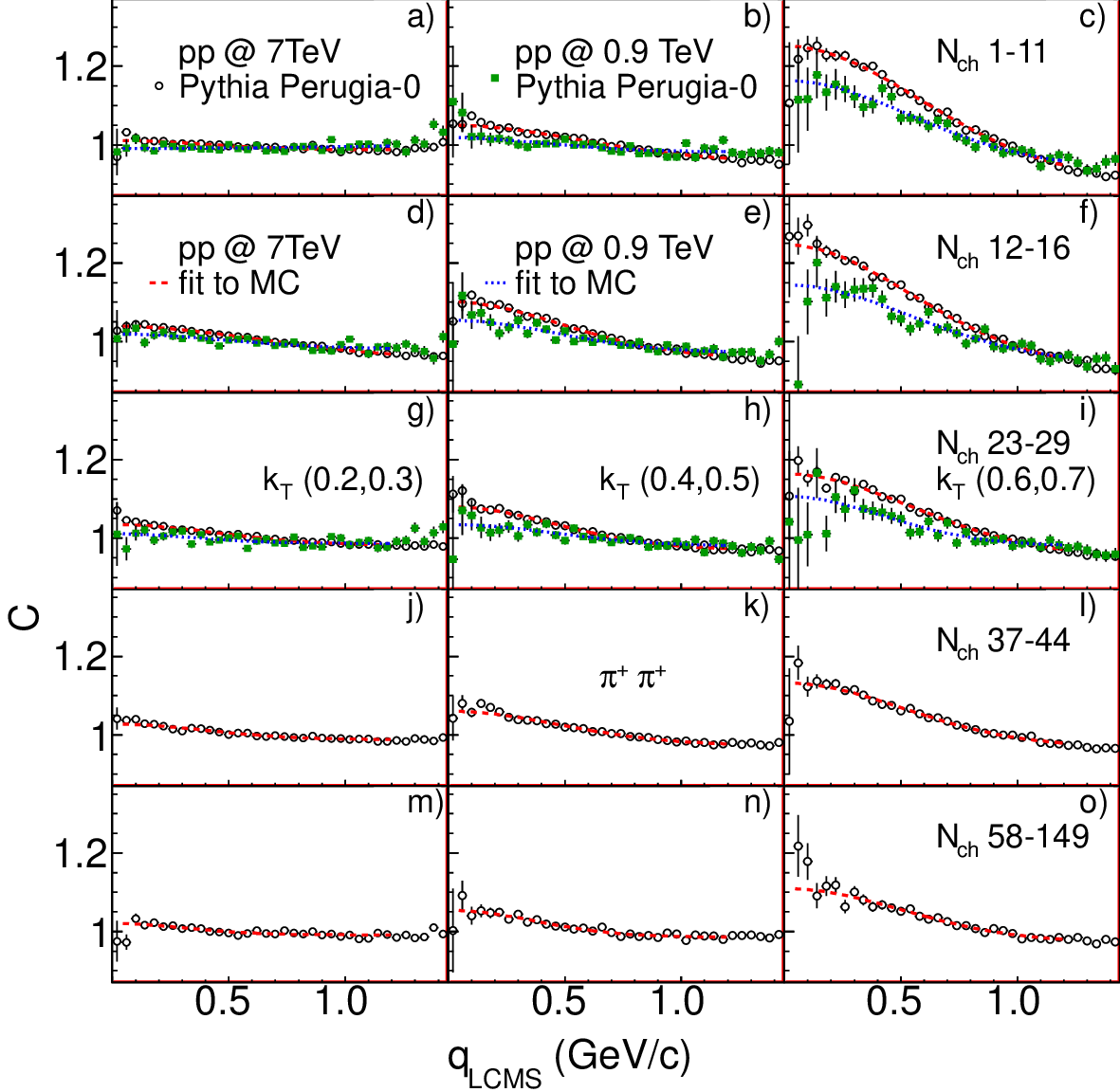}
}
\caption{\label{fig:mcback}
  Summary of the MC simulations for selected multiplicity and $k_{\rm T}$
  intervals, open symbols are a simulation at $\sqrt{s} = 7$~TeV,
  closed symbols at $\sqrt{s} = 0.9$~TeV. Dashed lines are Gaussian
  fit to the simulations to determine the background parameters (see
  text for details).  
}
\end{figure*}
  
In order to characterize the non-femtoscopic background we study in
detail the correlation structure in the MC generators, in exactly the
same multiplicity/$k_{\rm T}$ ranges as used for data analysis. We see
trends that are consistent with the ``mini-jet'' hypothesis. The
correlation is small or non-existent for low $p_{\rm T}$ (first
$k_{\rm T}$ range) and it grows strongly  with $p_{\rm T}$. In
Fig.~\ref{fig:mcback} we show this structure for selected
multiplicity/$k_{\rm T}$ at both energies. At the highest $k_{\rm
 T}$ the effect has the magnitude of $0.3$ at low $q$, comparable
to the height of the femtoscopic peak. The appearance of these
correlations is the main limiting factor in the analysis of the
$k_{\rm T}$ dependence. We tried to analyze the correlations at
$k_{\rm T}$ higher than $0.7$~GeV/$c$ but we were unable to obtain a
meaningful femtoscopic result, because the ``mini-jet'' structure was
dominating the correlation. The strength of the correlation decreases
with growing multiplicity (as expected), slower than $1/M$, so that it
is still significant at the highest multiplicity. We studied
other tunes of the {\sc pythia} model and found that the Perugia-0 tune
reproduces the ``mini-jet'' correlation structures best, which is why
it is our  choice. Its limitation though is a relatively small
multiplicity reach, smaller than the one observed in data. As a result
the MC calculation for our highest multiplicity range is less reliable
-- this is reflected in the systematic error. 

\begin{figure*}[]
\centerline{
\includegraphics[width=0.80\textwidth]{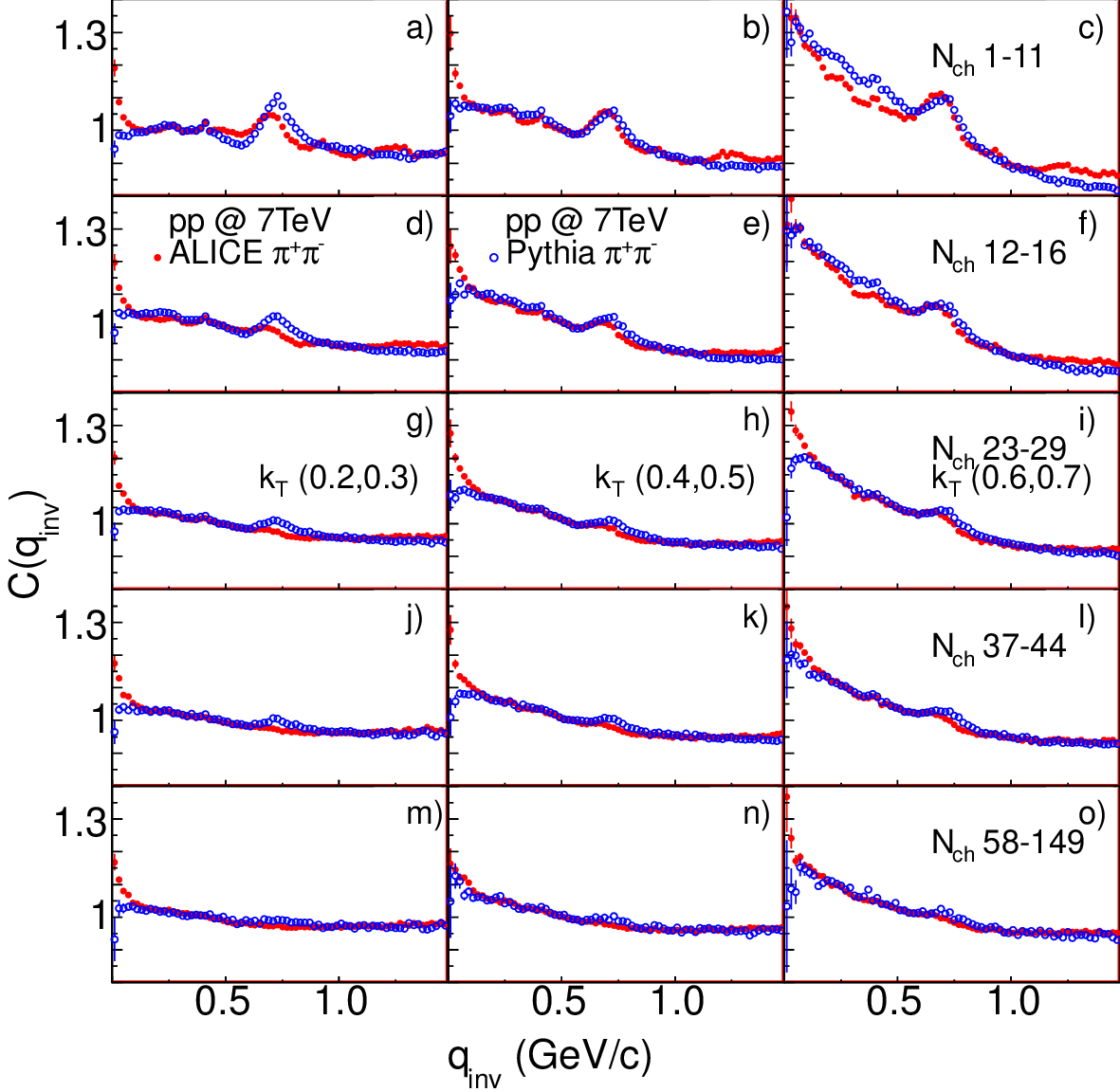}
}
\caption{\label{fig:mctodatapippimcf}
  Comparison of the correlation functions for $\pi^{+}\pi^{-}$ pairs
  at $\sqrt{s} = 7$~TeV (closed symbols) to the {\sc pythia} MC simulations
  (open symbols), in selected multiplicity and $k_{\rm T}$
  intervals. The plot is made as a function of $q_{inv}$ instead of
  $q_{LCMS}$ so that the resonance peaks are better visible.
}
\end{figure*}

Analyzing the shape of the underlying event correlation for identical
particle pairs in MC is important; however, it does not ensure that
the behavior of the correlation at very low $q$ is reproduced well in
MC. We compared the identical particle MC and data in the large
$q$ region, where the femtoscopic effect is expected to disappear, and
found them to be very similar in all multiplicity/$k_{\rm T}$. However,
if the ``mini-jet'' hypothesis is correct, the same phenomenon causes
similar correlations to appear for non-identical pions. The magnitude
is expected to be higher than for identical pions, because it is
easier to produce an oppositely-charged pair from a fragmenting
``mini-jet'' than it is to create an identically-charged pair, due to
local charge conservation. Moreover, the femtoscopic effect for
non-identical pions comes from the Coulomb interaction only. It is
limited to very low $q$, below $0.1$~GeV/$c$. It is therefore possible
to test the low-$q$ 
behavior of the ``mini-jet'' correlation with such correlations. In
Fig.~\ref{fig:mctodatapippimcf} we show the measured $\pi^{+}\pi^{-}$
correlation functions, in selected multiplicity/$k_{\rm T}$ ranges,
compared to the corresponding correlations from the same MC sample
which  was used to produce correlations in Fig.~\ref{fig:mcback}. The 
underlying event long-range 
correlation is well reproduced in the MC. We see some deviation in
the lowest multiplicity range, which is taken into account in the
systematic error estimation. At larger multiplicities the strength of
the correlation is well reproduced. By comparing the 3D function in SH
we checked that the shape in 3D $q$ space is also  
in agreement between data and MC. The magnitude for 
non-identical pions is slightly bigger than for identical pions, as
expected. The femtoscopic Coulomb effect at $q<0.1$~GeV/$c$ is also
visible. Another strong effect, even dominating at low multiplicity, 
are the peaks produced by the correlated pairs of pions coming from
strong resonance decays. They do appear in the MC as well, but they
are shifted and have different magnitude. This is the effect 
of the simplified treatment of resonance decays in {\sc pythia}, where phase
space and final state rescattering are not taken into account. By
analyzing some of the correlation functions in
Fig~\ref{fig:mctodatapippimcf} we were able to identify signals from
at least the following decays: two-body $\rho$, $f_0$, and $f_2$ mesons
decays, three-body $\omega$ meson decay, and also possibly $\eta$
meson two-body decay. Some residual $K^0_S$ weak decay pairs, which
are not removed by our DCA selection, can also be seen. All of these
contribute through the full $q$ range $(0.0,1.2)$~GeV/$c$. This fact,
in addition to the stronger ``mini-jet'' contribution to non-identical
(as compared to identical) correlations, makes the non-identical
correlation not suitable for the  
background estimation for identical pion pairs. We also note that
there appears to be very rich physics content in the analysis of
resonances decaying strongly in the $\pi^+\pi^-$ channel; however, we 
leave the investigation of this topic for separate studies. 

The study of the $\pi^+\pi^-$ correlations confirms that the 
MC generator of choice reproduces the underlying event
structures also at low $q$. We found that they are adequately
described by a Gaussian in LCMS for the $C_0^0$ component. The dashed
lines in Fig.~\ref{fig:mcback} show the fit of this form to
the correlation in MC. The results of this fit, taken bin-by-bin for
all multiplicity/$k_{\rm T}$ ranges, are the input to the fitting procedure
described in Section~\ref{sec:3dfits}. Similarly, the observed $C_2^0$
correlation can be characterized well by a Gaussian, with the
magnitude of $-0.01$ or less and a peak around $q=0.5$~GeV/$c$ with a
width of $0.25-0.5$~GeV/$c$. We 
proceed in the same way as for $C_0^0$; we fit the MC correlation
structures with this functional form and take the results as fixed
input parameters in the fitting of the measured correlations. 

\subsection{Fitting the correlation function}
\label{sec:3dfits}

Having qualitatively analyzed the correlation functions themselves we
move to the quantitative analysis. The femtoscopic part of the
correlation function is defined theoretically via the Koonin--Pratt
equation~\cite{Koonin:1977fh,Pratt:1984su}: 

\begin{equation}
\label{eq:koonin-pratt}
C(\vec q, \vec k) = \int S({\bf r}, \vec q, \vec k) \left |\Psi({\bf
    r},\vec q) \right
|^2 d^4 {\bf r} , 
\end{equation}
where $\vec q$ is the pair 3-momentum difference (the fourth component
is not independent for pairs of identical pions since masses of
particles are fixed), $\vec k$ is 
the pair total momentum, {\bf r} is the pair space-time separation at
the time when the second particle undergoes its last interaction,
$\Psi$ is the wave function of  
the pair, and $S$ is the pair separation distribution. The aim in the
quantitative analysis of the correlation function is to learn as much
as possible about $S$ from the analysis of the measured
$C$. The correlation function $C$ is, in the most general form, a
6-dimensional object. We reduce the dimensionality to 3 by 
factorizing out the pair momentum $k$. We do not study the dependence
on the longitudinal component of $k$ in this work. The dependence on
the transverse component of $k$ is studied via the $k_{\rm T}$ binning,
introduced in Section~\ref{sec:alicedata}. We assume that $S$ is independent of $k$ inside
each of the $k_{\rm T}$ ranges. We also note that for identical pions the
emission function $S$ is a convolution of two identical single
particle emission functions $S_1$. 

In order to perform the integral in Eq.~\eqref{eq:koonin-pratt} we must
postulate the functional form of $S$ or $S_1$. We assume that $S$ does
not depend on $q$. The first analysis is performed with $S_1$ as a
3-dimensional ellipsoid with Gaussian density profile. This produces
$S$ which is also a Gaussian (with $\sigma$ larger by a factor of
$\sqrt{2}$): 

\begin{equation}
\label{eq:Sfun}
S(r_o, r_s, r_l) = N \exp \left (-\frac {r_o^2} {4 {R^G_{out}}^2}
  -\frac {r_s^2} {4 {R^G_{side}}^2} -\frac {r_l^2} {4 {R^G_{long}}^2}
\right ) , 
\end{equation} 
where $R^G_{out}$, $R^G_{side}$ and $R^G_{long}$ are pion femtoscopic
radii, also known as ``HBT radii'' or ``homogeneity lengths'', and
$r_{o}$, $r_{s}$ and $r_{l}$ are components of the pair separation
vector. For identical charged pions $\Psi$ should take into account
the proper symmetrization, as well as  
Coulomb and strong interaction in the final state. In the case of the
analysis shown in this work, with pions emitted from a region with the
expected size not larger than $2-3$~fm, the strong interaction
contribution is relatively small and can be
neglected~\cite{Lednicky:2005tb}. The influence of the Coulomb
interaction is  approximated with the Bowler--Sinyukov method. It
assumes that the Coulomb part can be factorized out from $\Psi$ and
integrated independently. 
There are well-known limitations to this approximation but they have
minor influence for the analysis shown in this work. 
With these assumptions $\Psi$ is a sum of two plane waves
modified by a proper  symmetrization. By putting Eq.~\eqref{eq:Sfun}
into Eq.~\eqref{eq:koonin-pratt} the integration can be done
analytically and yields the quantum statistics-only correlation
$C_{qs}$:  

\begin{equation}
\label{eq:Cfun}
C_{qs} = 1+\lambda \exp(-{R^G_{out}}^2 q_{out}^2 -{R^G_{side}}^2 q_{side}^2 -{R^G_{long}}^2q_{long}^2),
\end{equation}
where $\lambda$ is the fraction of correlated pairs for which both
pions were correctly identified. The 3-dimensional correlation
function is then modified with the Bowler--Sinyukov formula to obtain
the complete femtoscopic component of the correlation $C_{f}$:  

\begin{eqnarray}
\label{eq:bertsch-pratt}
C_f(\vec q)&=&(1-\lambda) + \lambda K(q_{inv})\\ 
& \times & \left [1+
 {\rm
   exp}(-{R^G_{out}}^2q_{out}^2-{R^G_{side}}^2q_{side}^2-{R^G_{long}}^2q_{long}^2) 
\right ],  \nonumber
\end{eqnarray}
where $K$ is the Coulomb like-sign pion pair wave function squared 
averaged over the Gaussian source with a radius of
$1$~fm. Changing this radius within the range of values measured in
this work has negligible effect on the extracted radii.
Eq.~\eqref{eq:bertsch-pratt} describes properly the femtoscopic  part
of the two pion correlation function. However, in the previous section
we have shown that our experimental functions also contain  
other, non-femtoscopic correlations. We studied them
in all multiplicity/$k_{\rm T}$ ranges and found that they can be
generally described by a combination of an angle-averaged Gaussian in
LCMS plus a small Gaussian deviation in the $C_2^0$ component:

\begin{eqnarray}
B(\vec{q}_{LCMS}) &=& A_h \exp(-|\vec{q}_{LCMS}|^2 A_w^2)  \\
&+& B_h \exp \left (\frac{-(|\vec{q}_{LCMS}|-B_m)^2} {2B^2_w} \right
) (3\cos^2(\theta) - 1), \nonumber
\label{eq:Bfun}
\end{eqnarray}
where $A_h$, $A_w$, $B_h$, $B_m$ and $B_w$ are parameters. They are
obtained, bin-by-bin, from the fit to the MC simulated correlation
functions shown in Fig.~\ref{fig:mcback}. They are fixed in the
procedure of fitting the data. The final functional form that is used
for fitting is then: 

\begin{equation}
C(q_{out},q_{side},q_{long}) = N C_f(q_{out},q_{side},q_{long}) B(q_{out}, q_{side}, q_{long}) ,
\label{eq:FitFinal}
\end{equation}
where $N$ is the overall normalization. Projections of the Cartesian
representation of the correlation functions, shown in
Figs.~\ref{fig:projectionsct} and~\ref{fig:ktdepcfct}, are normalized
with this factor. Function~\eqref{eq:FitFinal} is used to fit both the
SH and Cartesian representation of the 3D correlations.   

\begin{figure}[t!]
\centerline{
\includegraphics[width=0.4\textwidth]{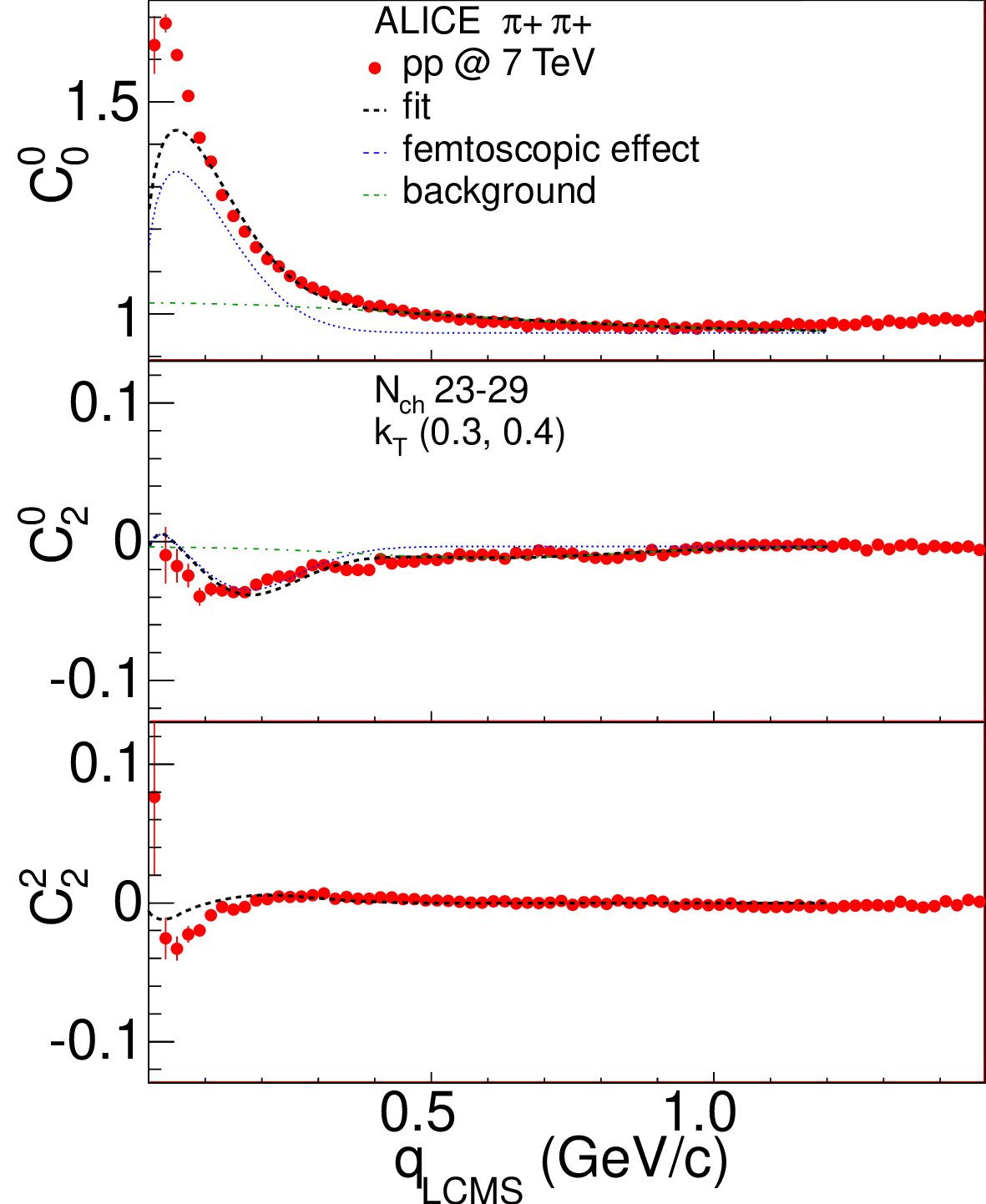}
}
\caption{\label{fig:FitExGaus}
Moments of the SH decomposition of the correlation functions for
events with $23\leq N_{ch} \leq29$ and pairs with $0.3<k_{\rm
  T}<0.4$~GeV/$c$. The dashed line shows the Gaussian fit, the
dash-dotted line shows the background component, the dotted line show
the femtoscopic component.
}
\end{figure}

\begin{figure}[t!]
\centerline{
\includegraphics[width=0.4\textwidth]{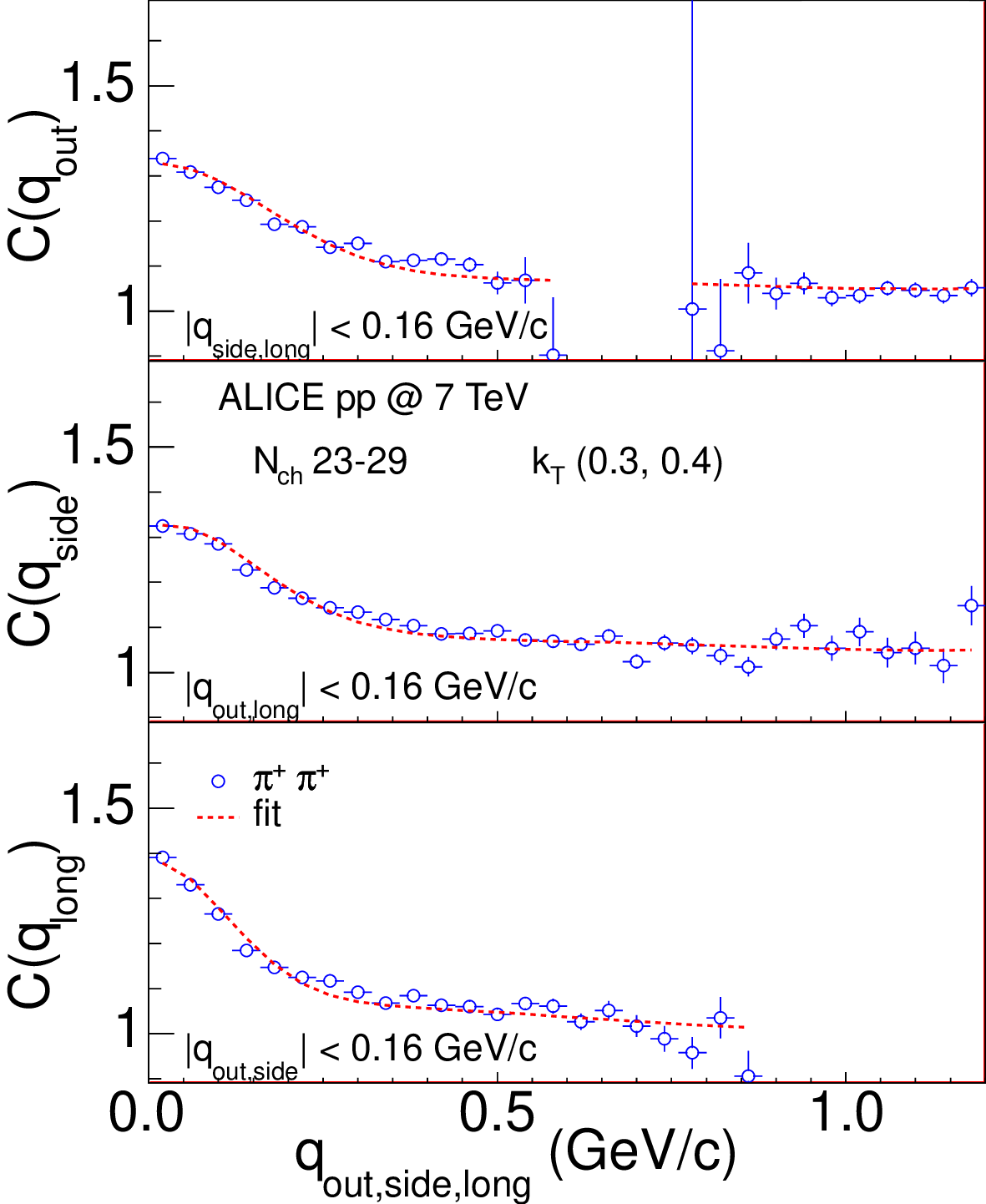}
}
\caption{\label{fig:FitExGausCT}
Projections of the 3D Cartesian representations of the correlation
functions for events with $23\leq N_{ch}\leq29$ and pairs with $0.3<k_{\rm
  T}<0.4$~GeV/$c$. To project onto one $q$-component, the others are
integrated over the range $0-0.16$~GeV/$c$. Dashed lines show
analogous projections of the Gaussian fit.
}
\end{figure}

In Fig.~\ref{fig:FitExGaus} an example of the fit to one of our
correlation functions is shown. The SH representation
of the data is shown as points; the result of the fit is a black
dashed line. The femtoscopic component is shown as a blue dotted line,
the non-femtoscopic background as green dash-dotted line. The
correlation function in this range has significant contribution from the
background and is reasonably reproduced by the fit. At $q<0.1$~GeV/$c$
the fit misses the data points in $C_0^0$ and $C_2^2$; we will discuss
what can be done to improve the agreement in
Section~\ref{sec:non-gaussian}. In Fig.~\ref{fig:FitExGausCT} the same
correlation is shown as projections of the 3D Cartesian 
representation. The other $q$ components are integrated over the range
of $0-0.16$~GeV/$c$. The fit, shown as lines, is similarly projected. In
this plot the fit does not describe the shape of the correlation
perfectly; however, the width is reasonably reproduced. 

\section{Fit results}
\label{sec:fits}

\subsection{Results of the 3D Gaussian fits}
\label{sec:3dradii}

\begin{figure}[t!]
\centerline{
\includegraphics[width=0.46\textwidth]{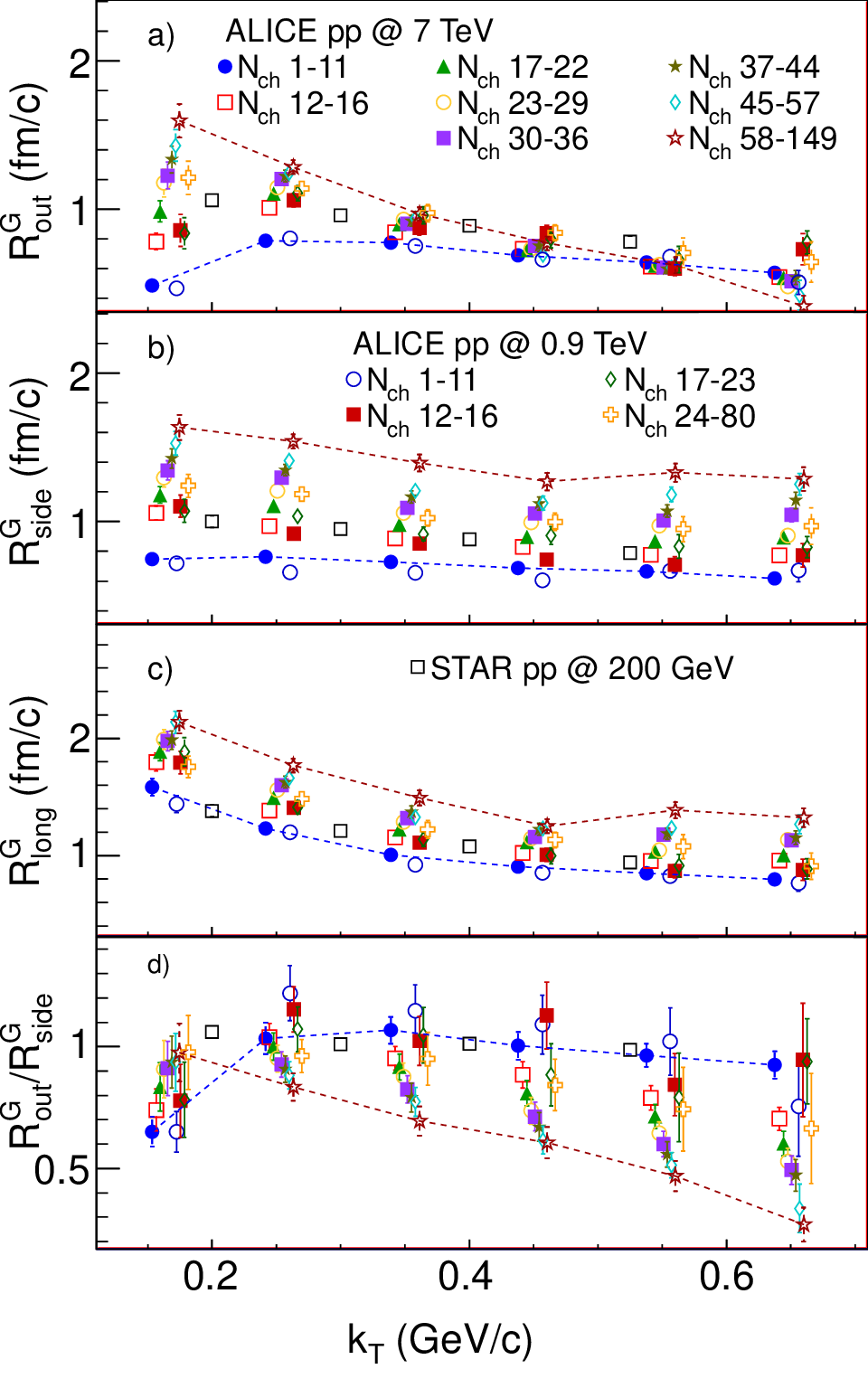}
}
\caption{\label{fig:RadiiSumGaus}
Parameters of the 3D Gaussian fits to the complete set of the
correlation functions in 8 ranges in multiplicity and 6 in $k_{\rm
  T}$ for \pp collisions at $\sqrt{s} = 7$~TeV, and 4 ranges in
multiplicity and 6 in $k_{\rm T}$ for \pp collisions at $\sqrt{s} =
0.9$~TeV. All points at given $k_{\rm T}$ bin should be at the same
value of $k_{T}$, but we shifted them to improve visibility. Open
black squares show values for \pp collisions at $\sqrt{s} = 200$~GeV
from STAR~\cite{STAR:2010bw}. Lines connecting the points for lowest
and highest multiplicity range were added to highlight the trends.
}
\end{figure}

We fitted all 72 correlation functions (4+8 multiplicity ranges for two
energies times 6 $k_{\rm T}$ ranges) with Eq.~\eqref{eq:FitFinal}. We
show the resulting femtoscopic radii in Fig.~\ref{fig:RadiiSumGaus} as
a function of $k_{\rm T}$. The strength of the correlation $\lambda$
is relatively independent of $k_{\rm T}$, is 0.55 for the lowest
multiplicity, decreases monotonically with multiplicity and reaches
the value of 0.42 for the highest multiplicity range.  The radii shown
in the Fig.~\ref{fig:RadiiSumGaus} are the main results of this
work. Let us now discuss many aspects of the data visible in this
figure.  

Firstly, the comparison between the radii for two energies, in the same
multiplicity/$k_{\rm T}$ ranges reveals that they are universally similar, at
all multiplicities, all $k_{\rm T}$'s and all directions. This confirms
what we have already seen directly in the measured correlation
functions. The comparison to $\sqrt{s} = 200$~GeV \pp collisions at
RHIC is complicated by the fact that these data are not available in
multiplicity ranges. The multiplicity reach at RHIC corresponds to a
combination of the first three multiplicity ranges in our study. No
strong change is seen between the RHIC and LHC energies. It shows that
the space-time characteristics of the soft particle production in \pp
collisions are only weakly dependent on collision energy in the range
between $0.9$~TeV to $7$~TeV, if viewed in narrow multiplicity/$k_{\rm
  T}$ ranges. Obviously the $\sqrt{s}=7$~TeV data have a higher
multiplicity reach, so the minimum-bias (multiplicity/$k_{\rm T}$
integrated) correlation function for the two energies is different.  

Secondly, we analyze the slope of the $k_{\rm T}$
dependence. $R^G_{long}$ falls with $k_{\rm T}$ at all multiplicities
and both energies. $R^G_{out}$ and $R^G_{side}$ show an interesting
behavior -- at low multiplicity the $k_{\rm T}$ dependence is flat for
$R^G_{side}$ and for $R^G_{out}$ it rises at low $k_{\rm T}$ and then
falls again. For higher multiplicities both transverse radii develop a
negative slope as multiplicity increases. At high multiplicity the
slope is bigger for $R^G_{out}$, while $R^G_{side}$ grows universally
at all $k_{\rm T}$'s while developing a smaller negative slope. The
difference in the evolution of shapes of $R^G_{out}$ and $R^G_{side}$
is best seen in their ratio, shown in panel d) of
Fig.~\ref{fig:RadiiSumGaus}. At low multiplicities the ratio is close
to 1.0, then it decreases monotonically with multiplicity. We note that a
negative slope in $R^G_{out}$ and $R^G_{side}$ was universally
observed in all heavy-ion measurements at RHIC energies and sometimes
also at lower energies. It is interpreted as a signature of a strong
collective behavior of matter created in such collisions. The
observation of the development, with increasing multiplicity, of such
slope in \pp collisions is consistent with the hypothesis, that the
larger the produced multiplicity, the more self-interacting and
collective is the produced system. 

\begin{figure}[t!]
  \centerline{\includegraphics[width=0.45\textwidth]{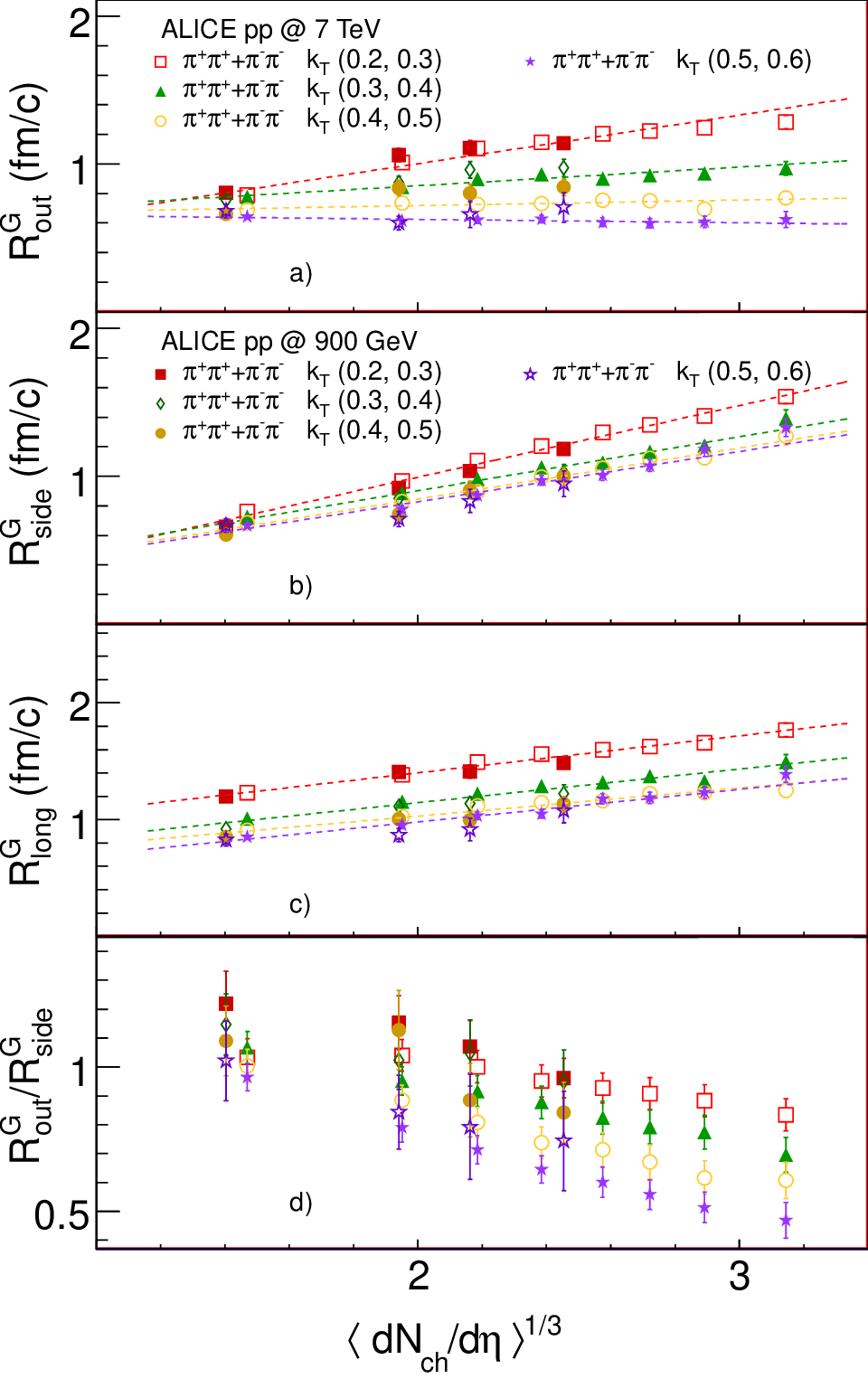}}
\caption{\label{fig:mdepALICEkt2}
Gaussian radii vs event multiplicity, for $\sqrt{s} = 0.9$~TeV and
$7$~TeV. Panel a) shows $R^G_{out}$, b) shows $R^G_{side}$, c) shows
$R^G_{long}$ and d) shows $R^G_{out}/R^G_{side}$ ratio. Lines show
linear fits to combined $\sqrt{s}=0.9$~TeV and $\sqrt{s}=7$~TeV points
in each $k_{\rm T}$ range. 
}
\end{figure}

Thirdly, all the measured radii grow with event multiplicity,
in each $k_{\rm T}$ range separately. This is shown more clearly in
Fig.~\ref{fig:mdepALICEkt2}, where we plot the radii as a function of 
$\left<{\rm d}N_{ch}/{\rm d}\eta\right>^{(1/3)}$ (For our \pp data we use the
$\left<{\rm d}N_{ch}/{\rm d}\eta\right>^{(1/3)}|_{N_{ch}\geq1}$ given in
Tab.~\ref{tab:mult}). $R^G_{side}$ and $R^G_{long}$ grow linearly with
the cube root of charged particle multiplicity, for all $k_{\rm T}$
ranges. Data, at both energies, follow the same scaling. For
$R^G_{out}$ the situation is similar for medium $k_{\rm T}$
ranges. The lowest $k_{\rm T}$ points show the strongest growth with
multiplicity, while the highest hardly grows at all. That is the
result of the strong change of the slope of $k_{\rm T}$ dependence
with multiplicity, noted in the discussion of
Fig.~\ref{fig:RadiiSumGaus}.  

\begin{figure}[t!]
  \centerline{\includegraphics[width=0.45\textwidth]{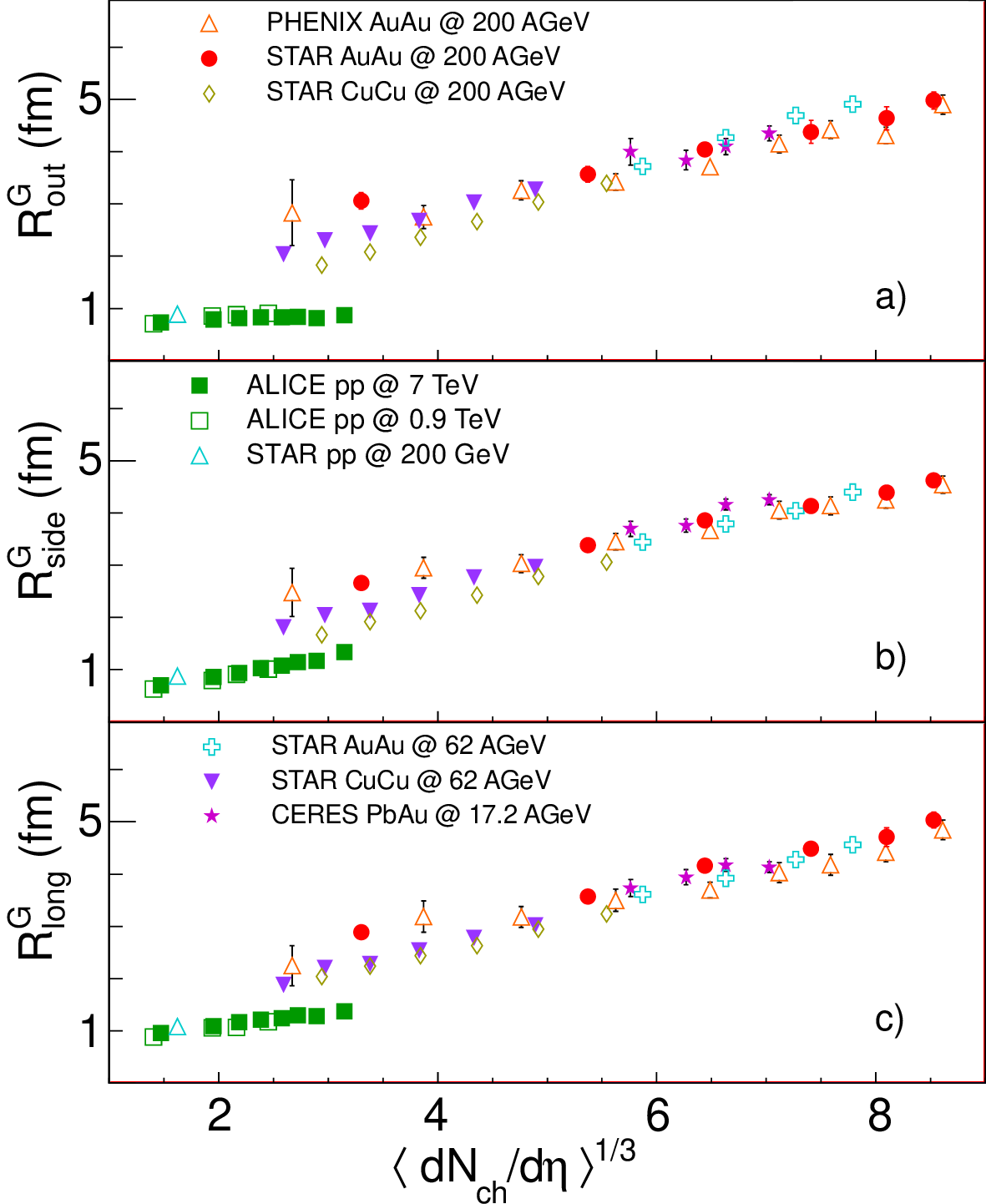}}
\caption{\label{fig:mdepWORLDkt2}
Gaussian radii as a function of $\left<{\rm d}N_{ch}/{\rm
    d}\eta\right>^{(1/3)}$, for $\sqrt{s} = 0.9$~TeV and $7$~TeV,
compared to the results from (heavy-)ion collisions at
RHIC~\cite{Adler:2004rq,Abelev:2009tp} and
SPS~\cite{Adamova:2002wi}. Panel a) shows $R^G_{out}$,  b) shows
$R^G_{side}$,  c) shows $R^G_{long}$. All results are for
$\left<k_{\rm T}\right>=0.4$~GeV/$c$, except the values from the
PHENIX experiment, which are at $\left<k_{\rm
    T}\right>=0.45$~GeV/$c$. 
}
\end{figure}

Similar multiplicity scaling was observed in heavy-ion collisions at
RHIC energies and below. In Fig.~\ref{fig:mdepWORLDkt2} we compare our
results to heavy-ion results from collision energies above
$15$~AGeV. This is the first time that one can directly compare \pp
and heavy-ion radii at the same $\left<{\rm d}N_{ch}/{\rm
    d}\eta\right>$, as  we measure $\left<{\rm d}N_{ch}/{\rm
    d}\eta\right>$ comparable to the one  in peripheral AuAu and CuCu
collisions at RHIC. Since the value of the radius strongly depends on
$k_{\rm T}$, we carefully selected the results to have the same
average  $\left<k_{\rm T}\right>=0.4$~GeV/$c$. The picture at other
$k_{\rm T}$'s is qualitatively similar. While both the heavy-ion  and
\pp data scale linearly with $\left<{\rm d}N_{ch}/{\rm
    d}\eta\right>^{(1/3)}$, the slope of the dependence is clearly
different, for all directions. The \pp results are systematically
below the heavy-ion ones at similar multiplicity; therefore, the
``universal'' multiplicity scaling~\cite{Lisa:2005dd} observed in
heavy-ion   collisions does not hold for \pp collisions at
$\sqrt{s}=0.9$ and $7$~TeV. The \pp radii do scale linearly with
multiplicity, but with a different slope.  

We speculate that the difference comes from a different way that the
two types of collisions arrive at similar multiplicity. To produce a
large number of particles in \pp collision one needs a particularly 
energetic elementary collision that produces a lot of soft
particles. The region where they are created is on the order of the
incoming proton size and the growth of the size with multiplicity
comes from further reinteraction between particles after they are
born. In contrast, in heavy-ion collision we have a combination of many
elementary nucleon scatterings, each of them producing a relatively low
multiplicity. But each scattering happens at a different space-time
point and it is the distribution of these points that mostly
determines the final observed size. In this picture, one would expect
the heavy-ion sizes to be larger than the ones observed in \pp at the
same multiplicity.

\subsection{Systematic uncertainty}
\label{sec:systematic}

The correlation function is, to the first order, independent of the
single particle acceptance and efficiency. We performed the
analysis independently for many samples of data, that
naturally had single particle efficiencies different by up to
5\%. We analyzed positive and negative pions separately, data at
two magnetic field polarities, data from three different month-long
``LHC periods'', each of them having a slightly different detector
setup. Two-particle correlations from all these analyzes were
consistent within statistical errors. 

We studied the effect of momentum resolution on the correlation peak
with the MC simulation of our detector. At this low $p_{\rm T}$, below
$1$~GeV/$c$, the momentum resolution for tracks reconstructed in the
TPC is below 1\%. This was confirmed by several methods,
including the reconstruction of tracks from cosmic rays, and
comparison of the reconstructed $K^0_S$ mass peak position with the
expected value. The smearing of single particle momenta does result in
the smearing of the correlation peak: it makes it appear smaller and
wider. We estimated that this changes the reconstructed radius by
1\% for the femtoscopic size of $1$~fm; the  effect grows to 4\% for
the size of $2$~fm, as it corresponds to a narrower correlation
peak. 

In contrast to single particle acceptance, the femtoscopic correlation
function is sensitive to the two-track reconstruction effects,
usually called ``splitting'' and ``merging''. The ``splitting'' occurs
when one track is mistakenly reconstructed as two. Both tracks have
then very close momenta. This results in a sharp correlation peak at
low relative momentum. We have seen such effects in the data and we 
took several steps to remove them. Firstly, the requirement that
the track is simultaneously reconstructed in the TPC and ITS decreases
splitting significantly.  In addition, each cluster in the TPC is
flagged as ``shared'' if it is used in the reconstruction of more than
one track. The split tracks tend to produce pairs which share most of 
their clusters; therefore, we removed pairs that share more
than 5\% of the TPC clusters. We also look for configurations
where a single track is split in two segments in the TPC, e.g. by the
TPC central membrane or a TPC sector boundary. Such segments should be
correctly connected in the tracking procedure to form a single track
if the detector calibration is perfect. However, in a few rare cases
this does not happen and a split track can appear. Such pairs would
consist of two tracks that have a relatively small number of TPC
clusters and they would rarely both have a cluster in the same TPC
padrow. Therefore, we count, for each pair, the number of times that
both tracks have a separate (non-shared) cluster in a TPC
padrow. Pairs for which this number is low are removed. Both
selections are applied in the same way to the signal and background
distributions. As a consequence, the fake low-momentum pairs from
splitting are almost completely removed, and the remaining ones are
concentrated in a very narrow relative momentum $q$ range,
corresponding essentially to the first correlation function bin. The
inclusion of this bin has a negligible effect on the fitting result;
hence, we do not assign any systematic error on the fitting values from
these procedures. 

Another two track effect is merging, where two distinct tracks are
reconstructed as one, due to finite detector space-point
resolution. The ALICE detector was specifically designed to cope with
the track densities expected in heavy-ion Pb+Pb collisions,
which are expected to be orders of magnitude higher than
the ones measured in \pp collisions. More specifically, the ITS
detector granularity as well as TPC tracking procedure, which allows
for cluster sharing between tracks, make merging unlikely. We  
confirmed with detailed MC simulation of our detector setup that
merging, if it appears at all, would only affect the correlation
function in the lowest $q$ bin, which means that it would not affect
the measured radii. 

In summary, the systematic uncertainty on the raw measurement, the
correlation functions itself, is small. 

\begin{table}[tb]
\caption{Systematic uncertainty coming from varying ``mini-jet''
  background height/width by 10\%/5\% up/down. 
  \label{tab:sysupdown}}
\begin{tabular}{lccc}
  \hline
  \hline
  $k_{\rm T}$ (GeV/$c$) & $\Delta R^G_{out}$ \% & $\Delta R^G_{side}$ \% & $\Delta
  R^G_{long}$ \% \\
  \hline
  (0.13, 0.2) & 4   & 1  & 2   \\
  (0.2, 0.3)  & 4   & 3  & 2   \\
  (0.3, 0.4)  & 4   & 3  & 2   \\
  (0.4, 0.5)  & 7   & 4  & 4   \\
  (0.5, 0.6)  & 9   & 4  & 4   \\
  (0.6, 0.7)  & 13  & 7  & 7   \\
  \hline
  \hline
\end{tabular}
\end{table}

The most significant systematic uncertainty on the extracted radii
comes from the fact that we rely on the MC simulation of the
``mini-jet'' underlying event correlations. We fix the parameters of
the $B$ function in Eq.~\eqref{eq:FitFinal} by fitting it to the
correlations obtained from the MC generated events. We  confirmed
with the analysis of the non-identical $\pi^{+}\pi^{-}$ pairs that our
Monte-Carlos of choice, the Perugia-0 tune of the {\sc pythia} 6 model, and
the {\sc phojet} model reproduce the height and the width of the
``mini-jet'' peak with an accuracy better than 10\%, except the first
multiplicity range where the differences go up to 20\% for the highest
$k_{\rm T}$ range. We performed the fits to the correlation function
varying the parameters $A_h$ and $B_h$ of the $B$ function by $\pm
10$\%, and $A_w$ by 5\%. The fit values for the case when
$A_h$, $B_h$ are decreased and $A_w$ is increased (corresponding to
smaller ``mini-jet'' correlations) are systematically below the
standard values. For larger ``mini-jet'' correlations they are
systematically above. The resulting relative systematic uncertainty on
all radii is given in Table~\ref{tab:sysupdown}. The error is
independent of multiplicity, except for the first and last multiplicity
ranges, where it is higher by 50\%. This error is fully correlated
between multiplicity/$k_{\rm T}$ ranges. 

\begin{table}[tb]
\caption{Systematic uncertainty coming from comparing the fit values
  with background obtained from {\sc phojet} and {\sc pythia} simulations.   
  \label{tab:syspytpho}}
\begin{tabular}{lccc}
  \hline
  \hline
  $k_{\rm T}$ (GeV/$c$) & $\Delta R^G_{out}$ \% & $\Delta R^G_{side}$ \% & $\Delta
  R^G_{long}$ \% \\
  \hline
  (0.13, 0.2) & 7   & 4  & 2  \\
  (0.2, 0.3)  & 1   & 1  & 4  \\
  (0.3, 0.4)  & 1   & 1  & 4  \\
  (0.4, 0.5)  & 7   & 2  & 4  \\
  (0.5, 0.6)  & 7   & 3  & 4  \\
  (0.6, 0.7)  & 10  & 6  & 7  \\
  \hline
  \hline
\end{tabular}
\end{table}

Independently, we  performed the fits with the {\sc phojet} generator
and fixed the parameters of $B$ from them. The difference in the final
fitted radii between {\sc pythia} and {\sc phojet} background is taken as another
component of the systematic error, shown in Table~\ref{tab:syspytpho}. 

\begin{table}[tb]
\caption{Systematic uncertainty coming from varying the maximum fit
  range.   
  \label{tab:sysrep}}
\begin{tabular}{lccc}
  \hline
  \hline
  $k_{\rm T}$ (GeV/$c$) & $\Delta R^G_{out}$ \% & $\Delta R^G_{side}$ \% & $\Delta
  R^G_{long}$ \% \\
  \hline
  (0.13, 0.2) & 3   & 2  & 1  \\
  (0.2, 0.3)  & 4   & 4  & 3  \\
  (0.3, 0.4)  & 7   & 5  & 3  \\
  (0.4, 0.5)  & 7   & 5  & 1  \\
  (0.5, 0.6)  & 7   & 4  & 3  \\
  (0.6, 0.7)  & 10  & 4  & 4  \\
  \hline
  \hline
\end{tabular}
\end{table}

Another effect, visible in Fig.~\ref{fig:FitExGaus}, is that
the traditional Gaussian functional form does not describe the shape
of the correlation perfectly. As a result, the extracted
radius depends on the range used in fitting. Generally, the larger the
fitting range, the smaller the radius. We  fixed our maximum
fitting range to $1.2$~GeV, which is be sufficient to cover all
correlation structures seen in data. We estimate that the remaining
systematic uncertainty coming from the fitting range is less than 5\%.

We  always performed all fits separately to correlations for
$\pi^+\pi^+$ and $\pi^-\pi^-$ pairs. They are expected to give the
same source size; therefore the difference between them is 
taken as an additional component of the systematic uncertainty.  

\begin{table}[tb]
\caption{Systematic uncertainty coming from comparing the fits to two
  independent 3D correlation function representations. 
  \label{tab:sysfit}}
\begin{tabular}{lccc}
  \hline
  \hline
  $k_{\rm T}$ (GeV/$c$) & $\Delta R^G_{out}$ \% & $\Delta R^G_{side}$ \% & $\Delta
  R^G_{long}$ \% \\
  \hline
  (0.13, 0.2) & 9   & 5  & 15  \\
  (0.2, 0.3)  & 9   & 7  & 7   \\
  (0.3, 0.4)  & 4   & 2  & 2   \\
  (0.4, 0.5)  & 6   & 2  & 4   \\
  (0.5, 0.6)  & 8   & 3  & 4   \\
  (0.6, 0.7)  & 18  & 6  & 12  \\
  \hline
  \hline
\end{tabular}
\end{table}

We used two independent representations of the 3D correlation
functions: the ``Cartesian'' one uses standard 3-dimensional histograms
to store the signal and the mixed background. The SH one uses sets of
1-dimensional histograms to store the SH 
components plus one 3D histogram to store the covariances between
them (see section~\ref{sec:representations} for more details). The
fitting of the two representations, even though it uses the same 
mathematical formula~\eqref{eq:FitFinal}, is different from the
technical point of view. The SH procedure is more robust against
holes in the acceptance~\cite{Kisiel:2009iw}, visible in our data,
e.g. in Fig.~\ref{fig:ktdepcf}. In an ideal case both procedures
should produce  identical fit results; therefore, we take the
difference between the radii obtained from the two procedures as an
estimate of the  systematic uncertainty incurred by the fitting
procedure itself. The error is shown in Table~\ref{tab:sysfit} as a
function of $k_{\rm T}$. The large error at low $k_{\rm T}$ is coming
from the fact that the two procedures are sensitive to the holes in
the acceptance in a different way. It reflects the experimental fact
that, in these $k_{\rm T}$ ranges, pairs in certain kinematic regions
are not measured; therefore, the 
femtoscopic radius cannot be obtained with better accuracy. In
the highest $k_{\rm T}$ range the ``mini-jet'' underlying correlation is
highest and broadest. If our simple phenomenological parametrization
of it does not perfectly describe its behavior in full 3D space, it
can affect differently a fit in Cartesian and SH
representations. 

In summary, the combined systematic error is 10\% for all $k_{\rm T}$
and multiplicity ranges except the ones at the lower and upper
edge. It is 20\% for the lowest and highest $k_{\rm T}$ and for the
lowest and highest multiplicity range at each collision energy. It is
also never smaller than $0.1$~fm. 

\subsection{Non-Gaussian fits}
\label{sec:non-gaussian}
 
In the discussion of Fig.~\ref{fig:FitExGaus} we note that the
measured correlation function is not perfectly reproduced by a 3D
Gaussian fit. In our previous work~\cite{Aamodt:2010jj} and in the
work of the CMS collaboration~\cite{Khachatryan:2010un} it was noted
that the shape of the 1-dimensional correlation in the Pair Rest Frame
is better described by an exponential shape. Also, model
studies~\cite{Humanic:2006ib} suggest that pion production at these
energies has large contribution from strongly decaying
resonances. This is confirmed by the observation of significant
resonance peaks in the $\pi^+\pi^-$ correlation functions, seen
e.g. in Fig.~\ref{fig:mctodatapippimcf}. Resonances decay after random
time governed by the exponential decay law, which transforms 
into an exponential shape in space via the pair velocity. By
definition pair velocity exists in the $out$ and $long$ direction, and
vanishes in $side$. It is then reasonable to attempt to fit the
correlation with a functional form other than a simple
Gaussian, at least for the $out$ and $long$ components.  

If we keep the assumption that the emission function factorizes into
the $out$, $side$ and $long$ directions, we can write a general form
of the pair emission function:

\begin{equation}
S({\bf r}) = S_o(r_o) S_s(r_s) S_l(r_l).
\label{eq:Sosl}
\end{equation}  
We can independently change each component. We stress, however, that
only for a Gaussian there is an analytically known correspondence
between the pair separation distribution $S$ and single particle
emission function $S_1$. Two commonly used forms of $S$ are
exponential and Lorentzian. They have the desired feature that the
integration in Eq.~\eqref{eq:koonin-pratt} can be analytically carried
out and produce a Lorentzian and exponential in $C$ respectively. In 
order to select the proper combination of functional forms we seek
guidance from models. They suggest that at least in the $out$ and
$long$ direction the emission function is not Gaussian, and in some
cases seems to be well described by a Lorentzian. We performed a study
of all 27 combinations of the fitting functions for selected
multiplicity/$k_{\rm T}$ ranges. We found that universally the $out$
correlation function was best described by an exponential,
corresponding to Lorentzian emission function, which agrees with model
expectations. In contrast, the $side$ direction is equally well
described by a Gaussian or a Lorentzian: we chose the former
because the Lorentzian correlation function would correspond to
exponential pair emission function with a sharp peak at 0. We deem
this unlikely, given that the models do not produce such shapes. In
$long$, the correlation function is not Gaussian; hence, we
chose the exponential shape in $C$ for the fit. In conclusion, we
postulate that the source has the following shape:

\begin{equation}
S({\bf r})= \frac{1}{r_o^2 + {R^E_{out}}^2} \exp \left(-\frac{r_s^2}{4
  {R^G_{side}}^2}\right ) \frac{1}{r_l^2 + {R^E_{long}}^2} ,
\label{eq:Sege}
\end{equation}
which corresponds to the following form of the femtoscopic part of the
correlation function formula:

\begin{equation}
C_f=1+\lambda \exp\left(-\sqrt{{R^E_{out}}^2 q_{out}^2}-{R^G_{side}}^2
q_{side}^2-\sqrt{{R^E_{long}}^2 q_{long}^2}\right) .
\label{eq:Cege}
\end{equation}

\begin{figure}[t!]
  \centerline{
\includegraphics[width=0.4\textwidth]{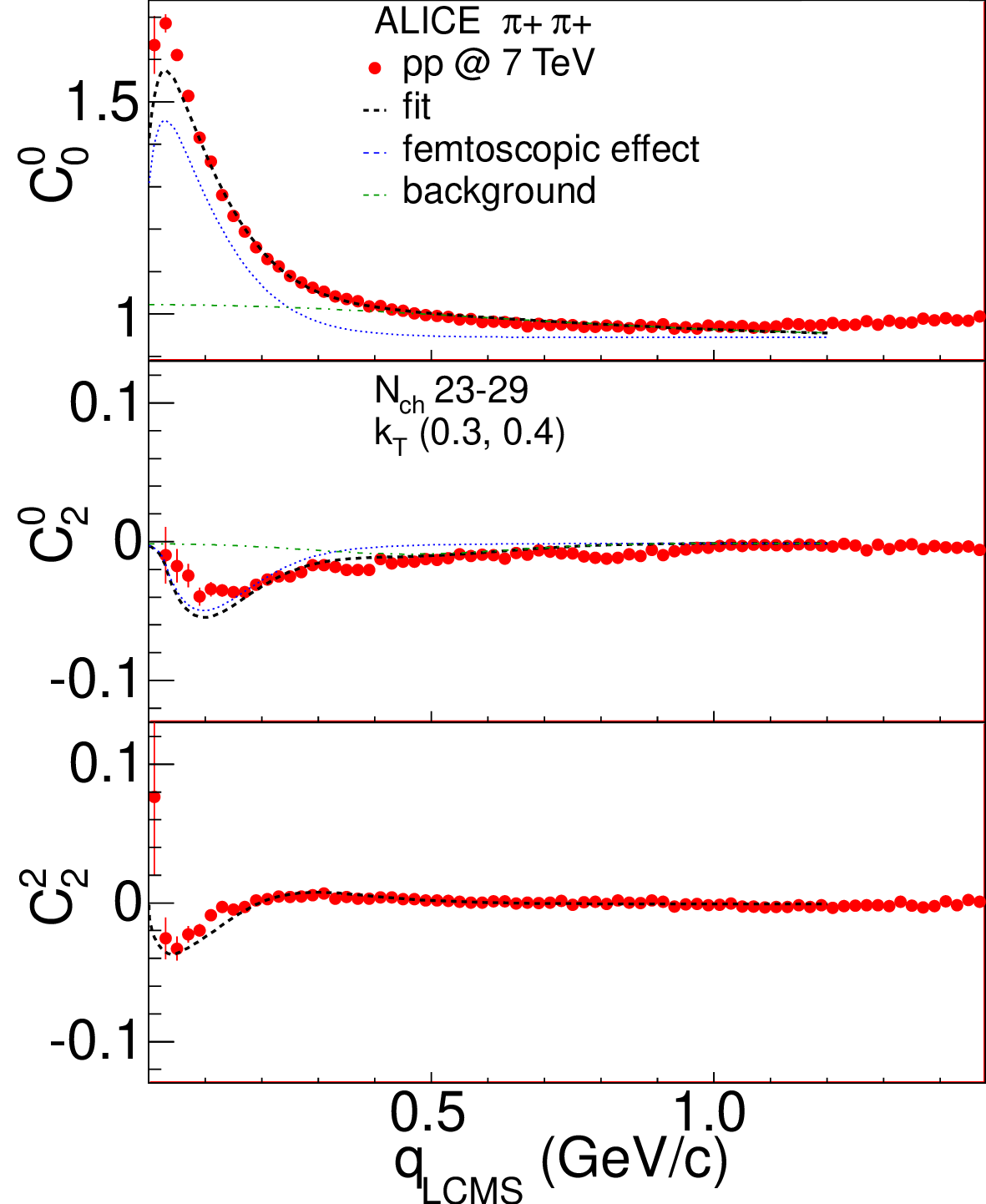}
} 
\caption{\label{fig:egefitexample}
  Exponential-Gaussian-exponential fit example for events with $23 \leq
  N_{ch} \leq 29$, pairs with $0.3 < k_{\rm T} < 0.4$~GeV/$c$. SH
  representation. Dotted line shows the femtoscopic component,
  dash-dotted shows the background, the dashed line shows the full fit.
}
\end{figure}

\begin{figure}[t!]
  \centerline{
\includegraphics[width=0.4\textwidth]{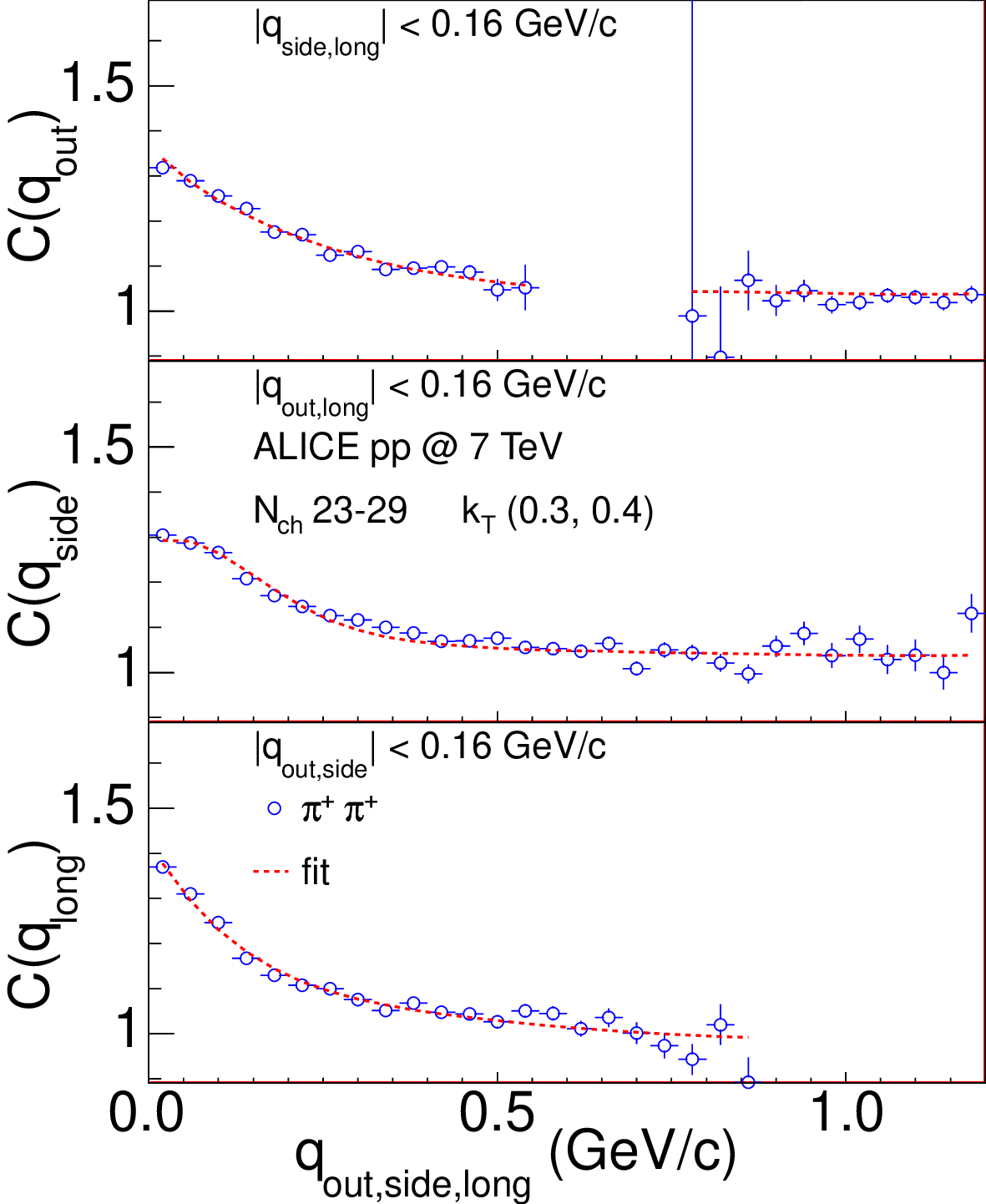}
} 
\caption{\label{fig:egefitexamplect}
  Exponential-Gaussian-exponential fit example for events with $23 \leq
  N_{ch} \leq 29$, pairs with $0.3 < k_{\rm T} < 0.4$~GeV/$c$. 1-dimensional 
  projections of the Cartesian representation are shown, the other $q$
  components were integrated in the range $0-0.16$~GeV/$c$.
} 
\end{figure}

In Figs.~\ref{fig:egefitexample} and~\ref{fig:egefitexamplect} we show
an example of the exponential-Gaussian-exponential fit to the
correlation functions at multiplicity $23\leq N_{ch}\leq29$ and \kt in
$(0.3,0.4)$~GeV/$c$. In the SH representation we see improvements over
the Gaussian fit from Fig.~\ref{fig:FitExGaus}. The behavior in
$C_0^0$ at low $q$ is now well described. In $C_2^2$ the ``wiggle'' in
the correlation is also reproduced -- this is possible because the 
functional forms for the $out$ and $side$ directions are now
different. In the Cartesian projections the improvement is also seen,
however it is not illustrated as clearly as in the SH.

\begin{figure}[t!]
  \centerline{
\includegraphics[width=0.48\textwidth]{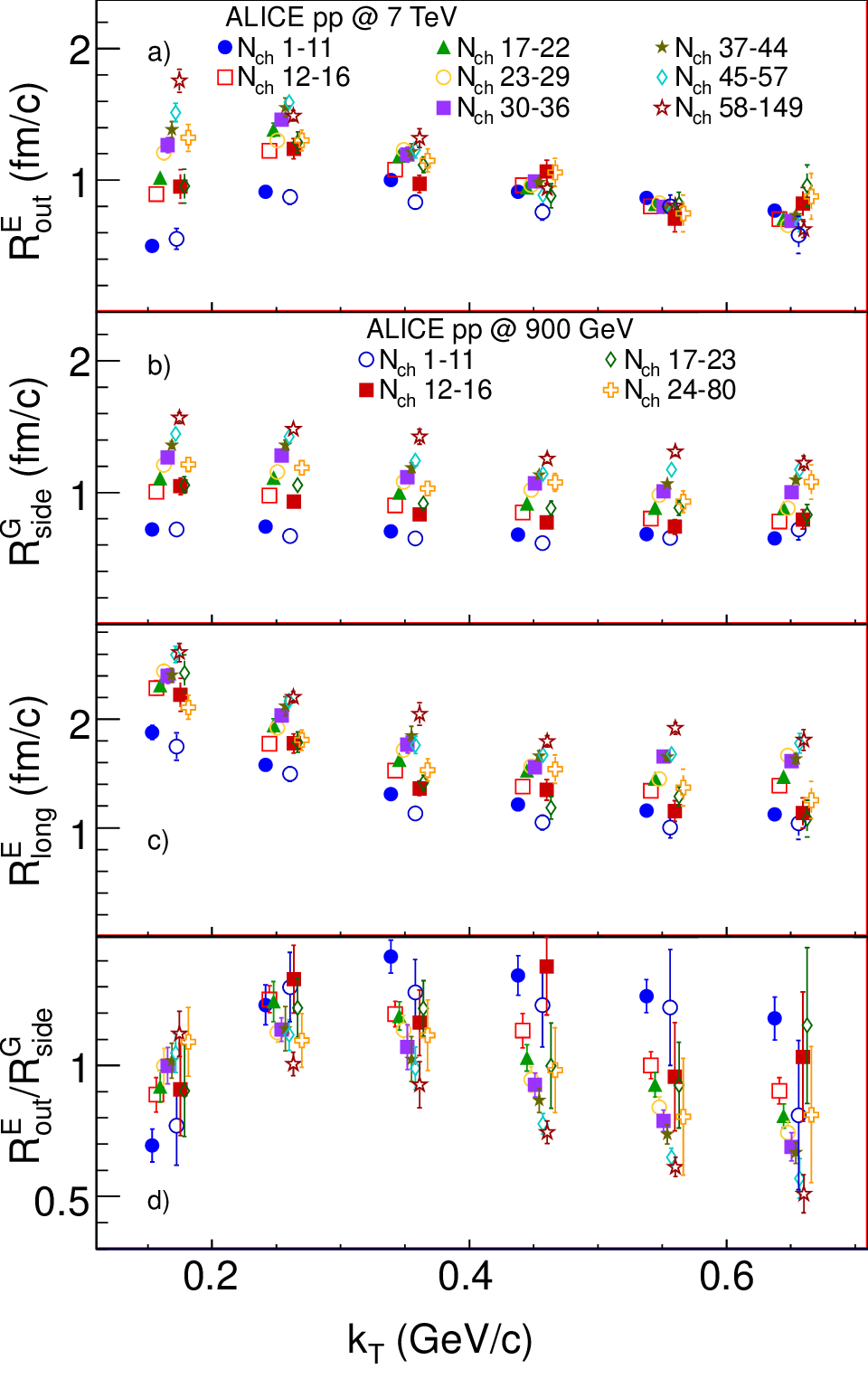}
}  
\caption{\label{fig:allradsege}
  Non-gaussian fit radii (see Eq.~\eqref{eq:Cege}) as a function of
  pair momentum $k_{\rm T}$ for all multiplicity ranges and for two
  collision energies. Panel a) shows $R^{E}_{out}$, panel b) shows
  $R^G_{side}$, panel c) shows   $R^E_{long}$, panel d) shows
  $R^E_{out}/R^G_{side}$ ratio. All points at given $k_{\rm T}$ bin
  should be at the same value of $k_{T}$, but we shifted them to
  improve visibility.  
}
\end{figure}

We then proceed with the fitting of the full set of 72 correlation
functions. The resulting fit parameters are summarized in
Fig.~\ref{fig:allradsege}. The quality of the fit (judged by the value
of $\chi^2/N_{dof}$) is better than for the 3D Gaussian fit. The
$\lambda$ parameter is higher by up to 0.2, as compared to the pure
Gaussian fit, reflecting the fact that the new functional form
accounts for the pairs contributing to the narrow correlation peak at
small $q$. The resulting exponential radii cannot be directly compared
in magnitude to the Gaussian radii from other experiments. However all
the features seen in dependencies of the Gaussian radii on
multiplicity and $k_T$ are also visible here. This confirms that with
a functional form that fits our correlation function  well (better
than a 3D Gaussian) the physics message from the dependence of radii
on multiplicity and $k_{\rm T}$ remains valid. The study of the fit
functional form shows that the correlation does not have a Gaussian
shape in $out$ and $long$.    

The $R^{G}_{side}$ from this fit should be equal to the $R^{G}_{side}$
from the 3D Gaussian fit with two caveats. The first is the assumption
that the emission function fully factorizes into separate functions for
$out$, $side$, and $long$ directions. In the fitting of the 3D
correlation functions the residual correlation between the value of
the $\lambda$ parameter and the values of the radii is often
observed. We noted already that the non-gaussian fit produces
larger values of $\lambda$, so $R^{G}_{side}$ could be
affected. Nevertheless we observe very good agreement (within
statistical errors for multiplicities above 16) between the
$R^{G}_{side}$ values from both fits, giving us additional confidence
that the underlying assumptions in our fit are valid.

Similar conclusions can be drawn from the ratio of the $R^E_{out}/R^G_{side}$
for the more advanced functional form, shown in panel d) of
Fig.~\ref{fig:allradsege}. Again, the picture seen for the Gaussian
radii is confirmed; the higher the multiplicity of the collision and
the collision energy, the lower the value of the ratio.

\begin{figure}[t!]
  \centerline{\includegraphics[width=0.4\textwidth]{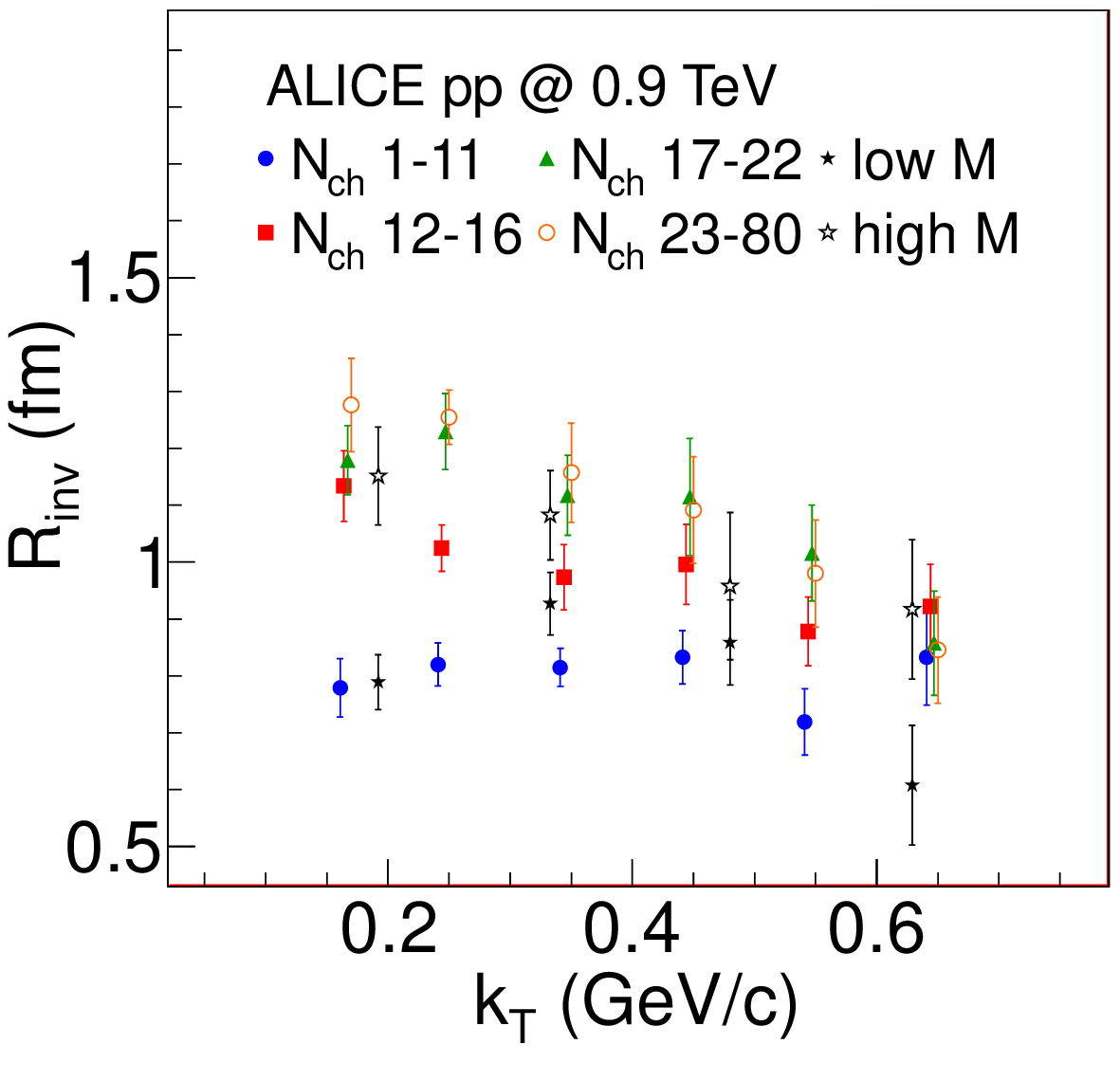}}
\caption{\label{fig:1drad900}
  1-dimensional $R_{inv}$ radius for all multiplicity and $k_{\rm
    T}$ ranges for the $\sqrt{s} = 0.9$~TeV data. The points for
  different multiplicities were slightly shifted in \kt for
  clarity. The systematic error, typically on the order of 10\% is not
  shown~\cite{Durham:2010aa}. Closed and open stars show the
  previously published result from~\cite{Aamodt:2010jj} for two ranges
  of the multiplicity $M$.   
}
\end{figure}
 
\section{Fitting 1D correlations}
\label{sec:1dfit}

For completeness, we also repeated the 1-dimensional study in Pair
Rest Frame, using all the methods and fitting functions described in
the previous work of ALICE~\cite{Aamodt:2010jj}. 
The 1-dimensional correlation functions are fit with the standard
Gaussian form, modified with the approximate Bowler--Sinyukov formula
to account for the Coulomb interaction between charged pions:   

\begin{equation}
\label{eq:1dbowler-sinyukov}
C(q_{inv})=\left [(1-\lambda)+
\lambda K(q_{inv}) \left (1 + {\rm
  exp}(-R_{inv}^2q_{inv}^2) \right) \right ] B(q_{inv}) ,
\end{equation}
where $K$ is the Coulomb function averaged over a spherical source
of the size $1.0$~fm, $R_{inv}$ is the femtoscopic radius and $B$ is
the function describing the non-femtoscopic background. In
Fig.~\ref{fig:1drad900} we plot the gaussian 1-dimensional invariant
radius as a function of multiplicity and $k_{\rm T}$. The closed and open
stars are the results from our earlier work, which are consistent with
the more precise results from this analysis. The systematic error is
on the order of 10\% and is now dominating the precision of the
measurement. At $\sqrt{s}=0.9$~TeV we see that, for the lowest
multiplicity, the radius is not falling with 
$k_{\rm T}$, while it develops a slope as one goes to higher
multiplicity. The 1-dimensional analysis is consistent with the 3D
measurement --- one needs to take into account that by going from LCMS
(3D measurement) to PRF (1D measurement) because it is necessary to
boost the $out$ radius by pair velocity, which is defined by $k_{\rm
  T}$. Then, one averages the radii in three directions to obtain the
1D $R_{inv}$. 

\begin{figure}[t!]
  \centerline{\includegraphics[width=0.4\textwidth]{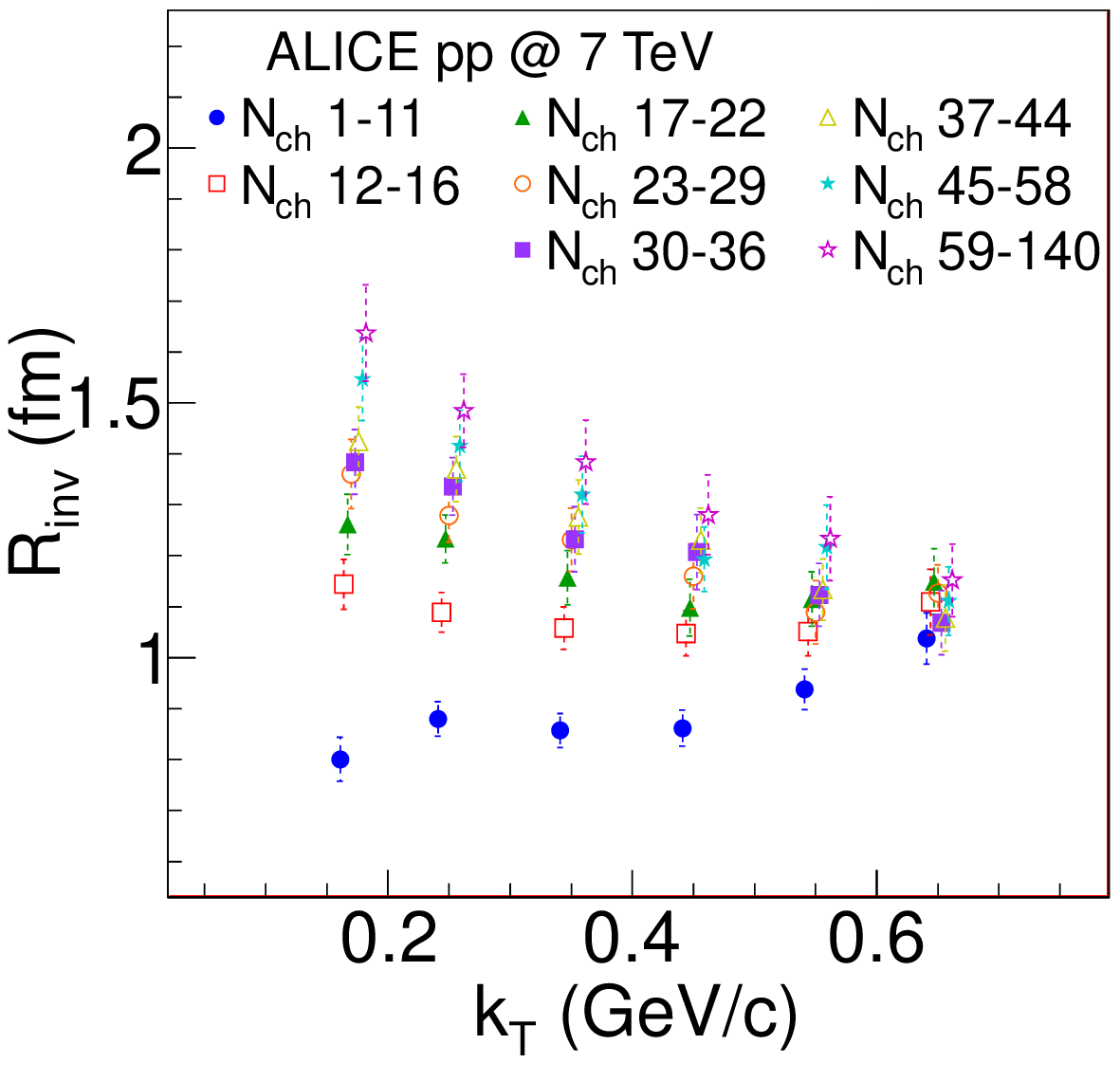}}
\caption{\label{fig:1drad7000}
1-dimensional $R_{inv}$ radius versus multiplicity and $k_{\rm T}$ for the
$\sqrt{s} = 7$~TeV data. The points for different multiplicities were
slightly shifted in \kt for clarity. The systematic error, typically
on the order of 10\% is not shown~\cite{Durham:2010aa}.  
}
\end{figure}

In Fig.~\ref{fig:1drad7000} we show the same analysis performed for the
$\sqrt{s}=7$~TeV data. The radii are again comparable at the same
multiplicity/$k_{\rm T}$ range. In addition, as one goes to higher
multiplicities, the $k_{\rm T}$ dependence of $R_{inv}$ is getting
more pronounced. The results are again consistent with the 3D
analysis. 

\section{Summary}
\label{sec:summary}

In summary, ALICE measured two-pion correlation
functions in \pp collisions at $\sqrt{s}=0.9$~TeV and at
$\sqrt{s}=7$~TeV at the LHC. The analysis was performed in
multiplicity and pair transverse momentum ranges. When viewed in the
same multiplicity and pair momentum range, correlation functions at
the two collision energies are similar.

The correlations are analyzed quantitatively by extracting the
emission source sizes in three dimensions: outward, sideward and
longitudinal. The longitudinal size shows expected behavior. It
decreases with pair momentum and increases with event multiplicity,
consistent with all previous measurements in elementary and heavy-ion
collisions. The transverse sizes show more complicated behavior. The
sideward radius grows with multiplicity and has a negative
correlation with pair momentum. The outward radius at the lowest
multiplicity is small for the lowest $k_{\rm T}$, increases for larger
$k_{\rm T}$ and then decreases. As the multiplicity grows the shape of the
$k_{\rm T}$ dependence gradually changes to the one monotonically
falling with $k_{\rm T}$. The 
resulting ratio of outward to sideward radii gets smaller as
multiplicity grows. Similar dependencies in heavy-ion collisions were
interpreted as signatures of the collective behavior of matter. One 
possible interpretation of the results in this work is that as one
moves towards \pp collisions producing high multiplicity of particles,
similar collectivity develops. More experimental and theoretical
information is needed to address this intriguing possibility. 

The upper range of multiplicities produced in \pp collisions at
$\sqrt{s}=7$~TeV is comparable to the multiplicities measured in
peripheral heavy-ion collisions at RHIC. When plotted  versus
$\left<{\rm d}N_{ch}/{\rm d}\eta\right>^{(1/3)}$ the radii in \pp show
linear  scaling, but with different slope and offset than those
observed in  heavy-ion collisions. Therefore our observations violate
the ``universal'' $\left<{\rm d}N_{ch}/{\rm d}\eta\right>^{(1/3)}$
scaling. This proves that the initial geometry of the collision does
influence the final measured radii and must be taken into account in
any scaling arguments.  

The analysis is complicated by the existence of the long-range
underlying event correlations. We assume these are the ``mini-jet''
structures which are visible at values of \pt as low as $0.5$~GeV/$c$. The
Monte-Carlo studies are consistent with such a hypothesis and are used
to parametrize and take into account the influence of mini-jets on the
fitted femtoscopic radii. Studies of the $\pi^{+}\pi^{-}$ correlations
are also consistent with such hypothesis. Nevertheless, the need to 
account for this effect remains the main source of the systematic
error.  

Finally, the detailed analysis of the correlation reveals that
the three-dimensional Gaussian describes the measurement only
approximately. A better shape, exponential-Gaussian-exponential, is
postulated, based on Monte-Carlo studies, and is found to better agree
with the data. The resulting radii and their behavior versus event
multiplicity and pair momentum are fully consistent with the one
obtained with the Gaussian approximation. 

\section{Acknowledgements}
\input{acknowledgements.tex}    %%%%%%% get the lates version before submitting

\bibliography{citations}
  
\end{document}

%% file: hbtpp-authors-revtex.tex
% Author list created by c:\users\kuijer\documents\visual studio 2010\projects\aliautorlist\aliautorlist\alixmldocument.cpp(Nov 21 2010)
% XML file: file:///C:/Documents and Settings/kuijer/Documents/Visual Studio 2010/Projects/AliAutorList/AliAutorList/authors-2010-12-12.xml
% XML created on: 2010-12-16T13:25:59
% merged with: 
%           file:///C:/Documents and Settings/kuijer/Documents/Visual Studio 2010/Projects/AliAutorList/AliAutorList/authors-exceptions-until2011.xml
%           file:///C:/Documents and Settings/kuijer/Documents/Visual Studio 2010/Projects/AliAutorList/AliAutorList/authors-exceptions-hbtpp.xml
%
\collaboration{ALICE} % CERN-LHC-ALICE
\noaffiliation

\author{K.~Aamodt}
\affiliation{Department of Physics and Technology, University of Bergen, Bergen, Norway}
\author{A.~Abrahantes~Quintana}
\affiliation{Centro de Aplicaciones Tecnol\'{o}gicas y Desarrollo Nuclear (CEADEN), Havana, Cuba}
\author{D.~Adamov\'{a}}
\affiliation{Nuclear Physics Institute, Academy of Sciences of the Czech Republic, \v{R}e\v{z} u Prahy, Czech Republic}
\author{A.M.~Adare}
\affiliation{Yale University, New Haven, Connecticut, United States}
\author{M.M.~Aggarwal}
\affiliation{Physics Department, Panjab University, Chandigarh, India}
\author{G.~Aglieri~Rinella}
\affiliation{European Organization for Nuclear Research (CERN), Geneva, Switzerland}
\author{A.G.~Agocs}
\affiliation{KFKI Research Institute for Particle and Nuclear Physics, Hungarian Academy of Sciences, Budapest, Hungary}
\author{S.~Aguilar~Salazar}
\affiliation{Instituto de F\'{\i}sica, Universidad Nacional Aut\'{o}noma de M\'{e}xico, Mexico City, Mexico}
\author{Z.~Ahammed}
\affiliation{Variable Energy Cyclotron Centre, Kolkata, India}
\author{N.~Ahmad}
\affiliation{Department of Physics Aligarh Muslim University, Aligarh, India}
% ...
% End of author list

%% file: acknowledgements.tex
The ALICE Collaboration would like to thank all its engineers and technicians for their invaluable contributions to the construction of the experiment and the CERN accelerator teams for the outstanding performance of the LHC complex.
The ALICE Collaboration acknowledges the following funding agencies for their support in building and
running the ALICE detector:
Calouste Gulbenkian Foundation from Lisbon and Swiss Fonds Kidagan, Armenia;
Conselho Nacional de Desenvolvimento Cient\'{\i}fico e Tecnol\'{o}gico (CNPq), Financiadora de Estudos e Projetos (FINEP),
Funda\c{c}\~{a}o de Amparo \`{a} Pesquisa do Estado de S\~{a}o Paulo (FAPESP);
National Natural Science Foundation of China (NSFC), the Chinese Ministry of Education (CMOE)
and the Ministry of Science and Technology of China (MSTC);
Ministry of Education and Youth of the Czech Republic;
Danish Natural Science Research Council, the Carlsberg Foundation and the Danish National Research Foundation;
The European Research Council under the European Community's Seventh Framework Programme;
Helsinki Institute of Physics and the Academy of Finland;
French CNRS-IN2P3, the `Region Pays de Loire', `Region Alsace', `Region Auvergne' and CEA, France;
German BMBF and the Helmholtz Association;
Greek Ministry of Research and Technology;
Hungarian OTKA and National Office for Research and Technology (NKTH);
Department of Atomic Energy and Department of Science and Technology of the Government of India;
Istituto Nazionale di Fisica Nucleare (INFN) of Italy;
MEXT Grant-in-Aid for Specially Promoted Research, Ja\-pan;
Joint Institute for Nuclear Research, Dubna;
 %
%Korea Foundation for International Cooperation of Science and Technology (KICOS);
National Research Foundation of Korea (NRF);
CONACYT, DGAPA, M\'{e}xico, ALFA-EC and the HELEN Program (High-Energy physics Latin-American--European Network);
Stichting voor Fundamenteel Onderzoek der Materie (FOM) and the Nederlandse Organisatie voor Wetenschappelijk Onderzoek (NWO), Netherlands;
Research Council of Norway (NFR);
Polish Ministry of Science and Higher Education;
National Authority for Scientific Research - NASR (Autoritatea Na\c{t}ional\u{a} pentru Cercetare \c{S}tiin\c{t}ific\u{a} - ANCS);
Federal Agency of Science of the Ministry of Education and Science of Russian Federation, International Science and
Technology Center, Russian Academy of Sciences, Russian Federal Agency of Atomic Energy, Russian Federal Agency for Science and Innovations and CERN-INTAS;
Ministry of Education of Slovakia;
CIEMAT, EELA, Ministerio de Educaci\'{o}n y Ciencia of Spain, Xunta de Galicia (Conseller\'{\i}a de Educaci\'{o}n),
CEA\-DEN, Cubaenerg\'{\i}a, Cuba, and IAEA (International Atomic Energy Agency);
The Ministry of Science and Technology and the National Research Foundation (NRF), South Africa;
Swedish Reseach Council (VR) and Knut $\&$ Alice Wallenberg Foundation (KAW);
Ukraine Ministry of Education and Science;
United Kingdom Science and Technology Facilities Council (STFC);
The United States Department of Energy, the United States National
Science Foundation, the State of Texas, and the State of Ohio.